\documentclass[nofootinbib,twocolumn,superscriptaddress,amsmath,amssymb]{revtex4}

\pdfoutput=1
\newcommand{\bcaption}[2]{\caption[#1]{\textbf{#1}#2}}
\newcommand{\homedir}{.}
\usepackage{graphicx}
\usepackage{rotating}
\usepackage{multirow}
\usepackage{float}
\usepackage{verbatim}
\usepackage{hyperref}

\newcommand{\px}[0]{~\rm{px}}
\newcommand{\micron}[0]{~\rm{\mu m}}
\newcommand{\nm}[0]{~\rm{nm}}

\newcommand{\ms}{~\rm{ms}}

\newcommand\given[1][]{\:#1\vert\:}
\newcommand{\by}[0]{\,\rm{x}\,}
\newcommand*\diff{\mathop{}\!\mathrm{d}}
\newcommand\tightapprox[0]{\approx\,\!}
\newcommand\tighttimes[0]{\!\times\!}

\newcommand{\Heaviside}{\mathrm{H}}

\DeclareMathOperator{\sinc}{sinc}
\DeclareMathOperator{\erf}{erf}

\newcommand{\zscale}[0]{z_{\rm{scale}}}
\newcommand{\zint}[0]{z_{\rm{int}}}
\newcommand{\CRB}[0]{Cram\'er-Rao bound}
\newcommand{\Psd}{P_s(\delta)}

\def\spvecA#1;{\if;#1;\else #1\cr \expandafter \spvecA \fi}

\let\oldhat\hat

\renewcommand{\vec}[1]{\boldsymbol{\mathbf{#1}}}
\renewcommand{\hat}[1]{\oldhat{\boldsymbol{\mathbf{#1}}}}

\begin{document}

\title{Light Microscopy at Maximal Precision}

\author{Matthew Bierbaum}
\thanks{These two authors contributed equally to this work.}
\affiliation{Department of Physics, Cornell University, Ithaca, NY 14853}
\author{Brian D. Leahy}
\thanks{These two authors contributed equally to this work.}
\affiliation{Department of Physics, Cornell University, Ithaca, NY 14853}
\author{Alexander A. Alemi}
\thanks{Work done while at Cornell, currently affiliated with Google Inc.}

\affiliation{Department of Physics, Cornell University, Ithaca, NY 14853}
\author{Itai Cohen}
\affiliation{Department of Physics, Cornell University, Ithaca, NY 14853}
\author{James P. Sethna}
\affiliation{Department of Physics, Cornell University, Ithaca, NY 14853}
\date{\today}

\begin{abstract}
Microscopy is the workhorse of the physical and life sciences, producing crisp images of everything from atoms to cells well beyond the capabilities of the human eye. However, the analysis of these images is frequently little better than automated manual marking. Here, we revolutionize the analysis of microscopy images, extracting all the information theoretically contained in a complex microscope image. Using a generic, methodological approach, we extract the information by fitting experimental images with a detailed optical model of the microscope, a method we call Parameter Extraction from Reconstructing Images (PERI). As a proof of principle, we demonstrate this approach with a confocal image of colloidal spheres, improving measurements of particle positions and radii by 100x over current methods and attaining the maximum possible accuracy. With this unprecedented resolution, we measure nanometer-scale colloidal interactions in dense suspensions solely with light microscopy, a previously impossible feat. Our approach is generic and applicable to imaging methods from brightfield to electron microscopy, where we expect accuracies of $1\nm$ and $0.1~\textrm{pm}$, respectively.
\end{abstract}

\maketitle

\section{Introduction}

\begin{figure*} [th]

\includegraphics[width=1.0 \textwidth]{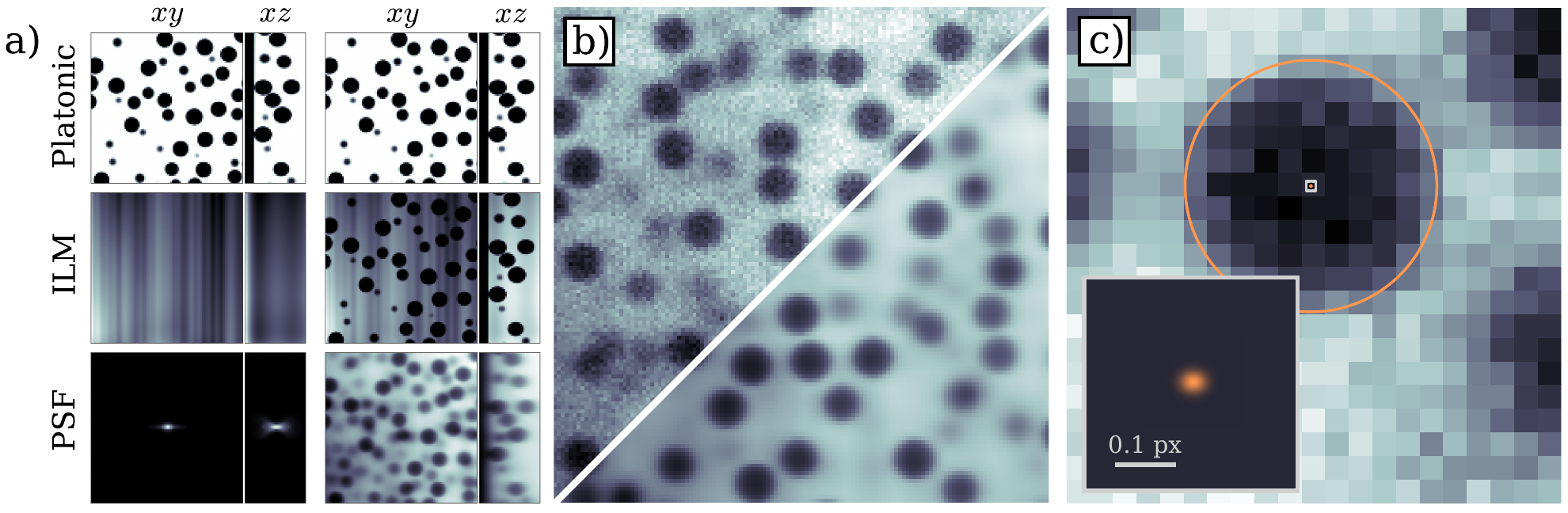}

\bcaption{PERI overview}{ -- A demonstration of model information recovered
from real confocal microscope images of $\langle a \rangle =
1.343(8)\micron$ colloidal spheres at a volume fraction of $\phi =
0.130(5)$. \textbf{(a)}: The generative model consists of a Platonic image of
dye distributed around perfect spheres and a coverslip (top), illuminated
with a spatially-varying intensity (middle), and convolved with a physical
point-spread function (bottom). The left panels show the model components,
and the right panels show the combined image. \textbf{(b)}: These components
combine to form a realistic generative model (bottom right), which aside from
noise is visually indistinguishable from the data (top left). \textbf{(c)}: From the
fit parameters of the model, we extract information such as particle positions
and radii (orange highlights) to within a few percent of a pixel -- corresponding
to an accuracy of $1$ and $3-4\nm$, respectively.
}

\label{fig:generative_model}
\end{figure*}

Microscope technology has progressed to near perfection. Crisp images speak of precisely engineered microscope components: large-aperture and nearly aberration-free lenses, high-frame-rate and low noise cameras, powerful and uniform light sources. Nanometer-scale details boast of super-resolution techniques thought impossible mere decades ago: PALM \cite{Betzig2006}, STORM \cite{Rust2006}, STED \cite{Hell1994}. The continued development of ever more powerful techniques -- SIM \cite{Kner2009}, Lattice-light sheet microscopy \cite{Chen2014} 
-- reassures that resolution will continue to improve.

However, our ability to extract \textit{quantitative} information from microscopy images has not kept pace. In fields from electron microscopy to super-resolution localization, current methods mimic human perception with heuristic approaches, such as looking for the centers of bright spots or regions of contrast in an image~\cite{Rogers2007, Parthasarathy2012, Grull2011, Anthony2009, Smith2010, Andersson2008}. The simplicity of these methods necessarily ignores physical complexities in the image formation. As a result, systematic errors and inefficient estimates plague these techniques~\cite{Gao2009, Lu2013}.

In this paper, we present a universal method of scientific image analysis that extracts \textit{all} the information theoretically contained in a complex image. Our method, dubbed parameter extraction from reconstructing images (PERI), uses a detailed model of the physics of image formation to fit experimental images. From the fit, we then extract information about the image at the information-theoretic limit. We illustrate this approach on confocal images of colloidal spheres, measuring each particle's position and radius to within $3\nm$, a 100x improvement over current methods. We use this extreme accuracy to measure colloidal interactions at the nanometer scale, measuring deviations from hard-sphere interactions for the first time with light microscopy. Our method does not require modifying the microscope or the image acquisition. As a result, any researcher with a microscope can readily apply our technique to push their data to the information-theoretic limit.

How precisely can an object be located in an image? The fundamental limitation in locating an object arises from statistical noise in the image formation, not directly from diffraction or optical limitations~\cite{Ram2006}. This limit is determined through the interplay of the image signal and noise, as described by the Cram\'er-Rao Bound. Specifically, the Cram\'er-Rao Bound states that the covariance matrix of the estimated parameters is always larger than the inverse of the Fisher information matrix of the noise distribution \cite{Rao1945}. For an image with Gaussian white noise of variance $\sigma^2$, sampled at points $\vec{x}_k$, the minimum uncertainty in the parameters $\vec{\theta}$ measured from the image is 

\begin{equation}\label{eq:crb_image}
\textrm{cov}\, \theta_{ij} \ge \sigma^2 \left( \sum_k \frac {\partial \mathcal{I}(\vec{x}_k)}{\partial \theta_i} \frac {\partial \mathcal{I}(\vec{x}_k)}{\partial \theta_j} \right)^{-1} \; ,
\end{equation}
where $\mathcal{I}(\vec{x})$ is the image that would be measured in the absence of noise.

We can use this equation to estimate the minimum uncertainty in measuring a colloidal sphere's radius and position from a three-dimensional confocal image. For a particle of radius $R$ blurred by diffraction over a width $w$, the derivatives with respect to particle radius in equation~(\ref{eq:crb_image}) are only nonzero on a shell at the particle's edge of approximately $4\pi R^2 w$ voxels. At the particle's edge, the intensity changes from a characteristic brightness $\tightapprox I$ to $\tightapprox 0$ over a width $\tightapprox w$, and the derivatives are thus of magnitude $\tightapprox I/w$. Substituting these values gives a minimum uncertainty in a particle's radius as $\sigma_R \sim \sqrt{w/4\pi R^2}/\textrm{SNR}$, where $\textrm{SNR}=I/\sigma$ is the signal-to-noise ratio. Likewise, changing the particle's position only affects the edge voxels in the direction of the particle's motion. The positional derivatives will thus be of magnitude $\tightapprox I/w$ only on a projected shell of $\approx\!\pi R^2 w$ voxels, giving the minimal uncertainty in the particle's position as $\sigma_x \sim \sqrt{w/\pi R^2}/\textrm{SNR}$. For a colloidal particle of diameter $1\micron$, imaged with a confocal microscope with voxel size of $100\nm$ and diffractive blur of $w\tightapprox200\nm$ at an $\textrm{SNR}=25$, these uncertainties correspond to $\sigma_R\!\approx\!1.5\nm$ and $\sigma_x\!\approx\! 3\nm$, a fantastically high precision. 

\section{Results}

Actually achieving this localization without serious systematic errors requires a detailed knowledge of the image formation process. To incorporate this knowledge, we create a generative model of the microscope image based on the physics of the light interacting with the sample and with the microscope's optical train. We then fit every parameter in the model by comparing the image produced by the model to the experimental image. Our model describes the physics of image formation in the order that it occurs: (1) fluorescent dye is distributed unevenly throughout the sample, (2) the dyed sample is illuminated unevenly by the laser, (3) the resultant image is blurred due to diffraction, and (4) the final image is noisy.

{\bf \emph{Dye Distribution:}} To reconstruct the image, we start with the continuous distribution of the fluorescent dye in the sample. For the image in Fig.~\ref{fig:generative_model}, the dye is distributed everywhere except in a slab, representing the glass cover-slip slide, and in a collection of spherical lacunae, representing the colloidal particles. To represent this continuous dye distribution on a pixelated grid, we draw these objects in real-space using a function that is tuned to match the exact Fourier representation of a sphere (see SI for an extensive discussion of this and the rest of the generative model). We call this correctly-aliased representation on a pixelated grid the Platonic image. While we focus on featuring only spheres in this work, PERI is flexible enough to include any parameterizable object in the generative model, such as ellipsoidal~\cite{Keville1991, Mohraz2005}, rodlike~\cite{Kuijk2011}, or polyhedral~\cite{Guttman2016} particles.

{\bf \emph{Illumination field and background:}} This distribution of dye is illuminated by a scanned laser. Due to imperfections and dirt in the optics, the illumination is not uniform but instead varies in space. For instance, our line-scanning confocal's illumination field is highly striped, as any imperfections in the line illumination are dragged across the field of view. We describe this spatially-varying illumination as a continuous field that varies throughout the image. Empirically, we find that combining a Barnes interpolant along the scan direction and Legendre polynomials in the perpendicular directions accurately describes both the rapidly-varying stripes and the slowly-varying changes in the illumination of our line-scan confocal. Additionally, the microscope always registers a non-zero background signal, which we include in our model. We parameterize this background similarly to the illumination field.

{\bf \emph{Point spread function:}} Diffraction prevents the illuminated dye from being imaged exactly onto the detector. Instead, each dye molecule in the sample projects a comparatively large blur, known as the point-spread function (PSF), onto the imaging camera. As a result, the image captured on the camera is a convolution of the illuminated Platonic image with the PSF, and not simply the illuminated dye itself. While complicated, this PSF has been calculated exactly by many researchers for different geometries~\cite{Hell1993, Visser1994, Zhang2007, Nasse2010, Conchello1994, Dusch2007, Wolleschensky2005, Botcherby2009}. For microscope samples with a refractive index different from what the optical train is designed for, the PSF worsens with depth, becoming significantly broader and more aberrated. We use an adaptation of these exact PSF calculations for a line-scanning confocal as our PSF model, optimizing over parameters such as the numerical aperture of the lens and the index mismatch of the sample to the optics.

Putting these components together as shown in Fig.~\ref{fig:generative_model}a, our model image $\mathcal{M}$ sampled at pixels $\vec{x}$ is described by
\begin{equation}
\mathcal{M}(\vec{x}) = B(\vec{x}) + \int \diff^3\vec{x}^{\prime}\,\,[I(\vec{x^{\prime}})(1-(1-c)\Pi(\vec{x^{\prime}}))] P(\vec{x}-\vec{x}^{\prime}; \vec{x})
\label{eq:genmodel}
\end{equation}
where $I$ is the illumination field, $B$ is the background, $\Pi$ is the platonic image, and $P$ is the spatially-varying PSF; we include a constant offset $c$ to partially capture rapidly-varying variations in the background. The model image is highly realistic, as shown by the comparison with real data in Fig.~\ref{fig:generative_model}b.

{\bf \emph{Noise:}} Finally, noise degrades the image recorded on the camera. We treat the noise using a Bayesian framework, and look for the maximum-likelihood model given the microscope data, complete with possible priors on parameter values. Since the noise is empirically Gaussian (see SI), the most likely model is the least-squares fit of the model to the microscope image.

To find the most likely model, we least-squares fit every parameter in our generative model to find the correct particle positions, radii, illumination field, and point-spread function, as illustrated in Fig.~\ref{fig:generative_model}c. A typical confocal image contains a few times $10^3$ particles, each with $4$ fit parameters ($x,y,z,R$). In addition, there are a few hundred global parameters to optimize, such as the illumination and PSF parameters and the lens's z-step size along the optical axis, resulting in $\tightapprox 10^4$ parameters per image -- a daunting optimization problem. We begin with an initial guess for the positions using standard particle locating techniques~\cite{Crocker1996}, and we simultaneously fit the particle positions and the global variables using a Levenberg-Marquardt algorithm modified for large parameter spaces \cite{Marquardt1963, Transtrum2010, Transtrum2011, Transtrum2012}. From here, we ensure that we have correctly identified every particle in the image by automatically adding and subtracting particles based on the the difference between the model and the microscope image. After finding the best-fit parameters, we sample from the log-likelihood using standard Monte Carlo techniques~\cite{Neal2003} to estimate the errors in the image reconstruction. (See SI for a detailed description of the fitting method and numerical optimizations.)

It is important to note that this fit is over all the pixels in the image -- to get a meaningful extraction of parameters, every pixel must be described accurately. Imperfectly fit regions -- due to e.g. deformed particles or PSF leakage from objects outside the image -- can bias the extracted positions of particles in the region and even affect the entire image reconstruction through the influence on image-scale variables.  

\begin{figure*}[th]

\includegraphics[width=0.67 \textwidth]{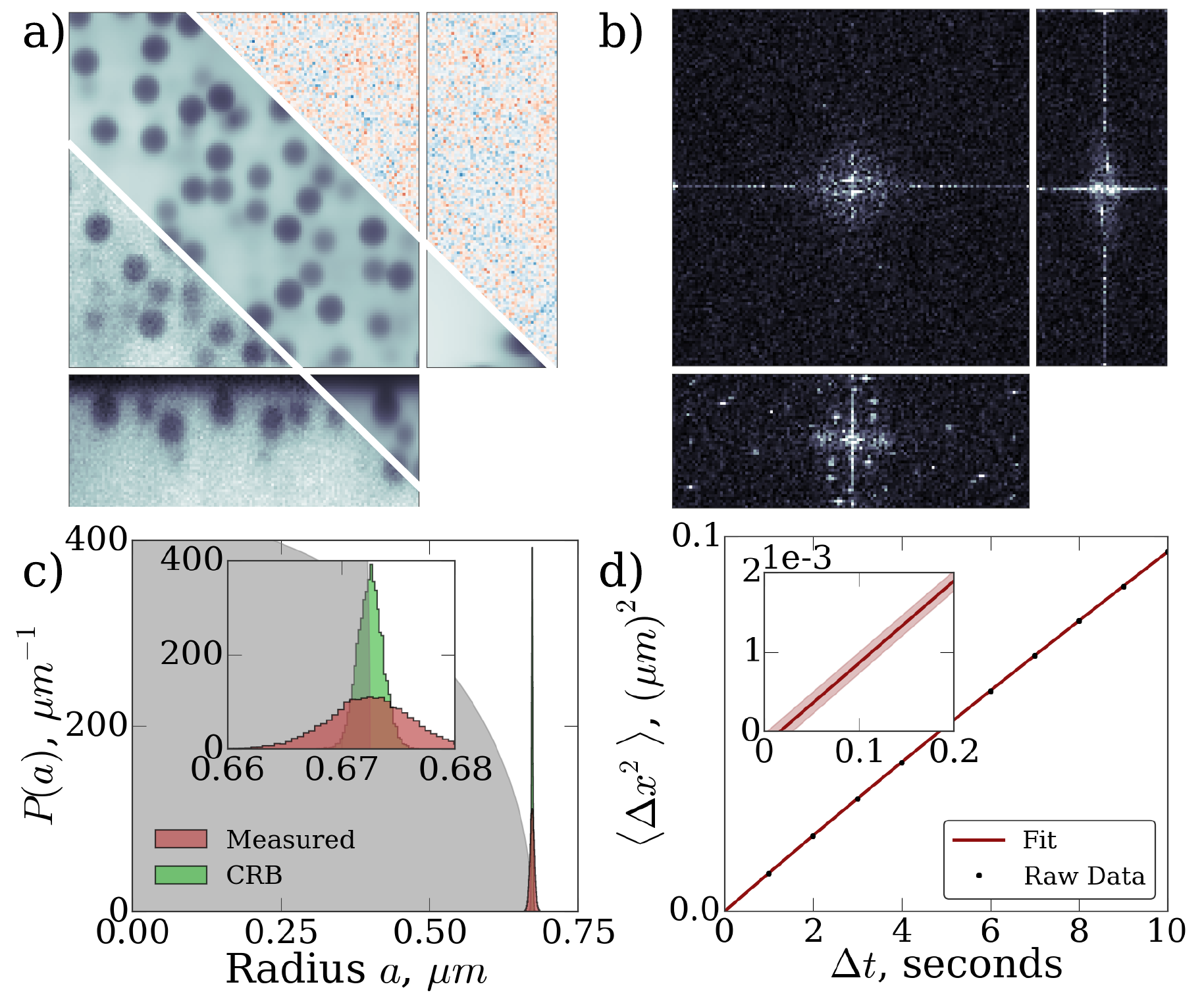}

\bcaption{Fitting the generative model to experimental data}{ --

\textbf{(a)} A representative image (lower left), its best-fit model (center), and the difference between
the two (upper right), shown as cross sections in the $xy$, $xz$, and $yz$ planes.
The residuals show nearly perfect Gaussian white noise at the expected signal-to-noise ratio,
demonstrating the quality of the generative model. \textbf{(b)} The Fourier power
spectrum of the same residuals, displayed as three orthogonal slices in the
$q_x q_y$, $q_x q_z$, and $q_y q_z$ planes. In addition to scanning noise,
visible as the stripes along $q_x=0$ and $q_z=0$ as well as the isolated
poles, excess power is visible at scales larger than the particles themselves
but smaller than the features given by the ILM. These residuals are associated
with the incomplete description of the point spread function. \textbf{(c)} PERI measures
the particle radius within an uncertainty of $3-4\nm$, as estimated from changes
in featured particle radii with time (red histogram). Improving the description
of the PSF would allow for radii to be featured at the CRB (green histogram), with a precision of $1\nm$. \textbf{(d)} The experimental
mean-squared displacement $\langle \Delta x^2\rangle$ (black dots; error bars smaller than symbol size) provides an estimate of PERI's average positional errors.
Extrapolating the fitted mean-square displacement (red curve; the shaded band denotes the fit uncertainty) to $t=0$ gives a positional error indistinguishable from $0\nm$.
}

\label{fig:experimental_data}
\end{figure*}

Using PERI to measure positions with nanometer accuracy requires rigorous checks on our method, with both generated and experimental data. We first generate images with a detailed physical model, employing an exact, spatially-varying point-spread function \cite{Hell1993}, experimentally-measured spatially-varying illumination, dense collections of particles with varying radii, and a realistic amount of noise. PERI successfully fits these generated data, converging to the global fit minimum in the extremely large dimensional parameter space despite a host of possible numerical complications, such as local minima in the fit space or a failure of the fit to converge. From this fit, PERI extracts both the particle positions and radii at the Cram\'er-Rao bound ($\tightapprox 2\nm$ and $\tightapprox 1\nm$, respectively). In contrast, current heuristic-based algorithms cannot measure the particle positions to better than 60~nm on realistically generated datasets. (See SI for a detailed comparison of PERI to other featuring algorithms.)

Emboldened by this success, we next test PERI on real experimental data. We take fast, three-dimensional movies of a suspension of $1.34 \micron$ diameter silica spheres suspended in a glycerol and water mixture and feature these images using PERI. By analyzing each frame in the movie independently, we can extract systematic errors from PERI's featuring.

We first analyze the residuals of our fits to the experimental data. Fig.~\ref{fig:experimental_data}(a,b) shows these residuals in both real- and Fourier-space. If our fit to the experimental image were perfect, the residuals would be perfectly Gaussian white noise. Instead, while the overall probability distribution of the residuals is nearly Gaussian in both domains (see SI), in Fourier-space there are distinct wave vectors above the noise floor. Comprising roughly $10^{-5}$ of the power in the experimental image, the extremely small size of this remaining signal demonstrates the quality of our generative model. The deviations of our model from the experimental data occur at length scales slightly larger than the particle diameter but smaller than typical illumination variations. These unexplained residuals most likely arise from approximations in models of line-scanning point spread function, excess aberrations in the microscope, and the artificially finite but large size we use in our PSF calculation to speed up optimization. Additionally, sharp peaks at high wave-vectors can be seen in one slice of the Fourier-space residuals, which arise from noise in the scanning of the lens and the line illumination. The remaining question is how much these residuals affect the parameters of interest, the particle positions and radii.

We can use the extracted particle positions and radii over time to test the accuracy of PERI. During the movies, the particles diffuse about, sampling different regions of the spatially-varying illumination and point-spread function and changing the configuration of neighboring particles. However, the true particle radii remain constant in time. Measuring individual radii fluctuations over time provides a stringent model-independent measurement of errors in PERI, as the changing configuration of the particles includes all the possible sources of systematic error. Tracking these radii fluctuations over time suggests that we can measure the particle radius to within 3-4$\nm$ (Fig.~\ref{fig:experimental_data}c), a fantastically high precision compared to the $672\nm$ particle radius and even the $125\nm$ pixel size. A better understanding of the image formation in the microscope could increase this precision even further, to the $1.5\nm$ minimal error from the Cram\'er-Rao Bound. We can also constrain the positional errors. Since the particle positions undergo Brownian motion, their mean-square displacement grows linearly in time $\langle \Delta x^2(t) \rangle = 2Dt$~\cite{Einstein1905}. Any featuring error that is uncorrelated with the particle position will manifest itself as a nonzero intercept when the fitted mean-square displacement is extrapolated to $t=0$. By extrapolating to zero (panel~d), we find that PERI's positional errors are indistinguishable from zero and are less than $10\nm$, with this constraint being limited only by statistics. Additionally, we check PERI on a dataset of 2$\micron$ diameter particles fixed in place via strong interactions -- a less demanding test since immobilizing the particles also fixes most of the sources of systematic error. In this data, we find $x$ and $y$ errors of 1-2$\nm$, $z$ errors of $3\nm$, and radii errors of $0.8\nm$ (see SI). Combined, these measurements demonstrate that we are able to measure particle positions and radii to within $3\nm$.

Why is PERI able to measure particle positions and radii so accurately while heuristic methods fail? Heuristic methods produce poor measurements with large systematic errors simply because they ignore complexities of the image formation, such as the spatially-varying illumination and  point-spread function. In contrast, PERI includes these complexities. Fitting the entire image ensures that all the complexities are accounted for -- any portion of the image formation not included in the model will manifest itself as strong residuals in the fit, declaring that the model is incomplete and suggesting what additional effect must be included. This process of model selection is described in detail in the SI.

This extraordinary accuracy in measuring particle positions from microscopy images creates a new window into nanometer-range particle interactions in dense suspensions. When colloidal particles are suspended in an aqueous solution, the particles charge, as the polar solvent dissociates ions on the particles' surface groups. This charge results in an electrostatic repulsion, which is in turn screened by counterions in the bulk~\cite{Russel1989, Israelachvili2011}. 
The screening creates an interparticle potential that deviates from a hard-sphere potential only at nanometer separations. This potential ever so slightly biases the distribution of particle positions away from that expected for a hard sphere suspension.

Previous efforts measured these interactions only in idealized, isolated surfaces such as a between two surfaces~\cite{Israelachvili1978} or a single colloidal particle interacting with a wall~\cite{Ducker1991, Prieve1999}. However, by their nature these idealized measurements frequently cannot include possible complications present in a real suspension, such as many-body interactions, realistic surface asperities, or increases in dissolved ion concentration from dissociated surface groups on multiple particles. Measuring the interaction potential in a dense colloidal suspension includes these and many other possible complications in the interaction.

\begin{figure*} [tp]

\includegraphics[width=1.0 \textwidth]{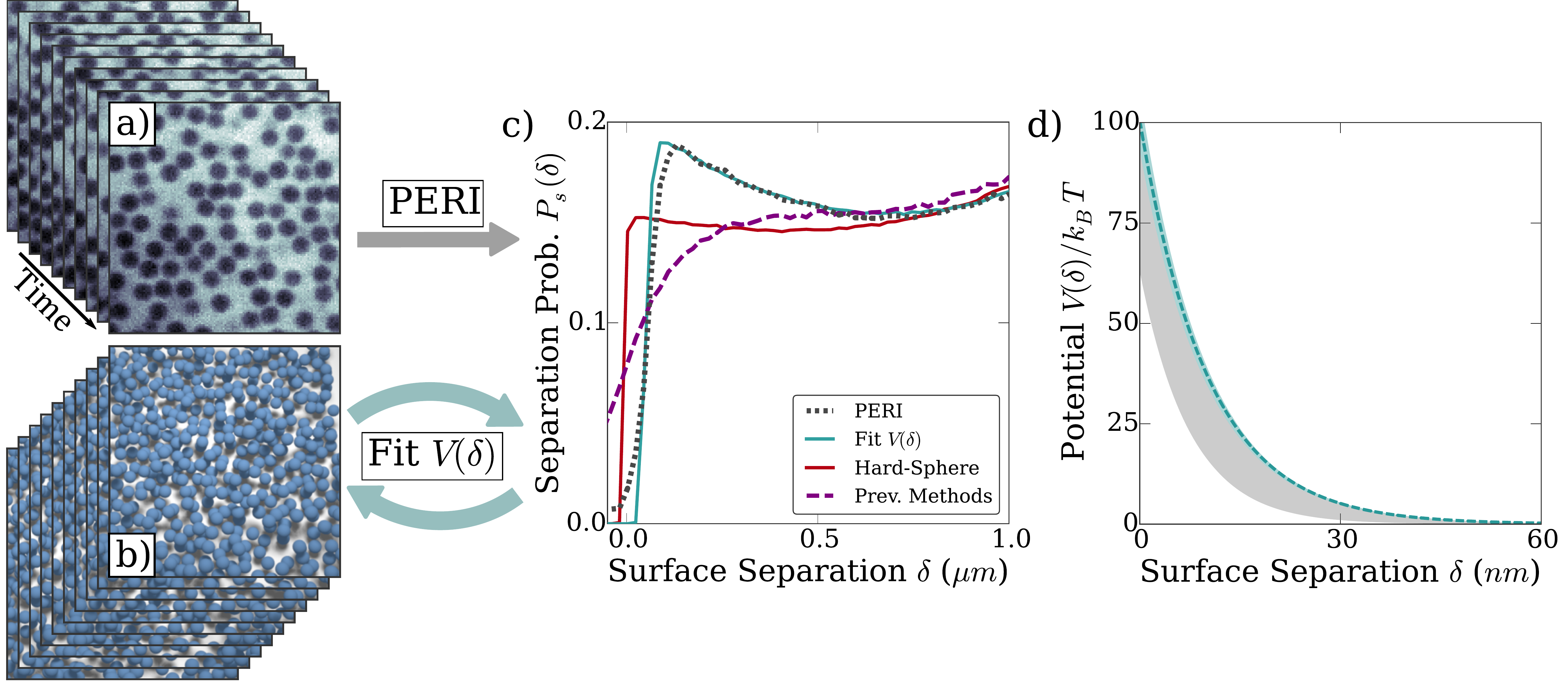}

\bcaption{Extracting Interparticle Potentials.}{
\textbf{(a)} We use PERI to analyze a large ensemble of three-dimensional images of a dilute suspension of $\approx$1200 Brownian particles; a small section of these images is shown in the upper left. From this data, we extract an experimental $\Psd$. \textbf{(b)} We use molecular dynamics
to create a simulated $\Psd$, and we iteratively update the interaction
potential $V(\delta)$ to find the $\Psd$ that best fits the experimental data.	\textbf{(c)} The extracted $\Psd$ from PERI (gray dashed line) and from the fitted
potential (solid cyan line) agree excellently. In contrast, the $\Psd$ from a strictly hard-
sphere potential (red line)
does not fit the data. The difference between these potentials depends on resolving particle
separations at the nanometer level. Previous centroid-based methods~\cite{trackpy} (purple line) produce a $\Psd$ with
nonsensical features, such as significant overlaps, that cannot be fit by a reasonable
interaction potential. \textbf{(d)} From the best-fit simulation, we extract the interparticle potential
$V(\delta)$. The shaded bands show the uncertainty in the potential, with the teal band
describing uncertainty in the fit and the gray band the uncertainty due to systematic errors (see SI for further discussion).
} 

\label{fig:exp}

\end{figure*}

We measure these nanometer-scale interactions by using PERI to analyze a large set of images of $1.3\micron$ silica spheres suspended in a water-glycerol mixture. To prevent kinetic effects from confounding our measurements, we allow the sample to fully sediment for an hour. This produces an open layer of sediment approximately 2-3 particle layers deep, shown in Fig.~\ref{fig:exp}a. We then image this suspension repeatedly over the course of several hours, extracting simulation-level detail of $\tightapprox$ 720,000 particle positions and radii over all the images. The particle interactions determine the structure of the suspension. We quantify this structure with the probability $\Psd$ of finding a pair of particles with surface-to-surface separation $\delta$, accounting for radii polydispersity and sedimentation in a manner preferable to the usual pair-correlation function. To reconstruct the interparticle potential, we use the extracted particle radii and particle number from the data and we simulate the particle dynamics using Brownian dynamics. We incorporate both gravitational settling and the interparticle potential, which we model as an exponentially-decaying electrostatic repulsion. We then fit the potential by simulating, reconstructing $\Psd$ from the simulation at each set of potential parameter values, and iterating to find the best $\Psd$ that matches experiment (Fig.~\ref{fig:exp}b).

The $\Psd$ from the best-fit simulation and from the experimental data analyzed by PERI agree excellently, at both large and small separations. At small separations, $\Psd$ rises rapidly over the first $\tightapprox 0.1\micron$ near contact in both the simulation data and the data extracted by PERI, as shown in figure~\ref{fig:exp}b. At longer distances (inset), the probability grows due to the increased volume where particles can be located, with slight oscillations reflecting second- and third- nearest-neighbor interactions. In contrast, previous centroid-based methods produce a $\Psd$ with nonsensical features, such as significant overlaps, that cannot be fit by a simulation.

We use the extracted $\Psd$ to measure nanometer-scale interactions in dense colloidal suspensions for the first time. The $\Psd$ measured by PERI is well-fit by an exponentially-decaying repulsive potential, as expected from electrostatic repulsion in standard colloidal theory~\cite{Russel1989} (figure~\ref{fig:exp}c). From the fit, we measure the potential's screening length as $10.1 \pm 2.5 \nm$ and the repulsion strength near contact as $100 \pm 30$ kT, corresponding to surface potentials and screening lengths similar to that previously measured from the interaction of a single particle with a wall~\cite{Ducker1991}. Our data strongly excludes hard-sphere interactions as the interparticle potential. Importantly, this resolving power between potentials results from the values of $\Psd$ near contact. Without the accurate localization provided by PERI, it is impossible to measure the potential at these separations.

\section{Discussion}

Our technique and the ideas within it provide more than just a description of colloidal interactions. Nanometer accuracy in locating colloidal particle positions would revolutionize fields as diverse as the study of colloidal glasses and the measurement of biological forces with force-traction microscopy. With our open-source code \footnote{Source code available with documentation and tutorials at \url{http://www.lassp.cornell.edu/sethna/peri/index.html}}, other researchers can immediately analyze existing images of these systems. Moreover, the principle of accurately reconstructing an image to extract parameters applies to a wide range of fields. Extending PERI to analyze brightfield microscopy images would provide nanometer-scale precision for a simpler and more widespread imaging setup than confocal microscopy. Applying these ideas to imaging modalities such as STEM or STM will usher in a new era of precision measurements, for objects whose sizes range from microns to angstroms.

\section{Materials and Methods}
﻿
The microscope is a Zeiss LSM 5 Live inverted confocal microscope, used in
conjunction with an infinity-corrected 100x immersion oil lens (Zeiss Plan-Apochromat, 1.4 NA, immersion oil with index $n=1.518$). The LSM 5 Live confocals operate by line-scanning. Rather than rastering a single point at a time to form the image, a line-scanning confocal images an entire line at once. An image of a line is focused onto the sample, and the sample fluorescence is detected on a line CCD. Rastering this line allows images to be collected extremely rapidly; the data in the text was taken at 108 in-plane frames per second. However, the different line-scanning optics worsen the point-spread function compared to a point-scanning confocal and cause illumination imperfections such as dirt to be smeared out over one direction in the image. Importantly, our confocal is outfitted with a hyper-fine piezo scanner which gives precise $z$-positioning of the lens. This precise $z$-positioning is important for accurate reconstruction of images -- with the less-precise standard positioning our image reconstruction and results suffer considerably.

Our experimental images consist of $\approx 1.3\micron$ silica particles
(MicroPearl) suspended in a mixture of glycerol and water. The glycerol/water
mixture is tuned to match the refractive index of the particles by minimizing
the sample scattering. For these particles we find the optimal refractive
index is $n \approx 1.437$ corresponding to $\approx 76\%$ glycerol and $24\%$
water. Since glycerol is hygroscopic, we controlled the concentration of glycerol and water by measuring the index of refraction rather than by measuring out the glycerol and water. We match the index of refraction of the spheres and the suspending fluid to within a few parts per thousand, resulting in practically zero scattering by the spheres of either the laser or fluorescent light. The glycerol has the additional advantage of creating a very viscous suspension, slowing down the Brownian motion of the particles. We add fluorescein sodium salt to dye the suspending fluid, at a concentration of 0.4 mg/mL. The fluorescein diffuses rapidly compared to the particles, and is effectively uniformly distributed throughout the regions occupied by the fluid. By using a considerable amount of dye and a low laser power, we minimize photobleaching during our experiments. Fluorescein sodium salt (molar weight 376.27) consists of two sodium ions bound to a dye molecule. Thus, this dye concentration corresponds to $\tightapprox 2 \times 10^{-3}$ moles/L of monovalent sodium ions and $10^{-3}$ moles/L of divalent fluorescein ions. 
To this solution we added the 1.3 $\micron$ silica particles (MicroPearl) at a concentration of 6.8 mg particles per 1 mL of solution. These particles are placed in a 100$\micron$ deep sample cell; since the particles sediment the experimental volume fraction is determined equally by settling and the sample cell height as opposed to simply the density of particles in the original suspension. We allow the suspension to sediment for several hours to achieve equilibrium before taking any measurements. The data is collected over the course of a 1-2 hours; we do not observe any change in the $\Psd$ from the earlier samples to the later ones.

\acknowledgments

We would like to acknowledge N. Lin, W. Zipfel, M. Transtrum, C. Clement, D. Koch, C. Schneider, and L. Bartell, S. Whitehead, T. Beatus and other members of the Cohen Lab for useful discussions. This work was supported in part by NSF DMR-1507607 (M. Bierbaum, A. Alemi, J. Sethna, and I. Cohen), NSF DMR-1120296 (B. Leahy), and ACS PRF 56046-ND7 (B. Leahy). This work used the Extreme Science and Engineering Discovery Environment (XSEDE), which is supported by National Science Foundation grant number ACI-1053575.

\bibliography{Full_Bibliography}

\pagebreak
\pagebreak
\onecolumngrid

\begin{center}
\textbf{\large Supplemental Material to: \\ Light Microscopy at Maximal Precision}
\end{center}

\setcounter{section}{0}
\setcounter{equation}{0}
\setcounter{figure}{0}
\setcounter{table}{0}
\setcounter{page}{1}
\makeatletter
\renewcommand{\theequation}{S\arabic{equation}}
\renewcommand{\thefigure}{S\arabic{figure}}
\renewcommand{\bibnumfmt}[1]{[S#1]}
\renewcommand{\citenumfont}[1]{S#1}

\renewcommand{\homedir}{./supplemental}
\section{Overview}

In this supplemental material we describe the details of our method for
extracting parameters from experimental confocal images at the highest
resolution possible without modifying the microscope itself.  To achieve
maximal resolution, we build a generative model which aims to describe the
value of every pixel in the experimental image.  That is, we create simulated
images by explicitly modeling every relevant aspect of image formation
including particle positions and sizes, the location of dirt in the optics,
amount of spherical aberration in the lens, and the functional form of the
point spread function.  We describe each of these model components in detail in
Section~\ref{sec:generative_model} and how we decided on these particular
components in Section~\ref{sec:model_selection}.  In order to fit this model to
the experiment, we adjust all model parameters until the features present in
the true experimental image are duplicated in the simulated one.  We decide
when the fit is complete and extract errors of the underlying parameters
by using a traditional Bayesian framework which is described in general terms
in Section~\ref{sec:bayesian_framework}.  This high dimensional optimization is
in general very difficult and so we describe our algorithmic improvements and
particular techniques in Section~\ref{sec:implementation}. Finally, we assess
the accuracy of this method in extracting underlying parameters and compare its
performance with traditional featuring methods in Section~\ref{sec:benchmarks}.

Overall, this document is meant to provide a roadmap for other researchers to
follow when adapting this technique to other types of microscopy and other
types of samples in order to extract the maximal amount of information from
their experimental images.

\section{Bayesian framework} \label{sec:bayesian_framework}

When fitting a model to noisy data, it is useful to adopt a Bayesian framework
in which we rigorously treat the noise as part of our model. In the case of
our featuring method, we fit a model of each image pixel $M_i$ to experimental
data $d_i$, which can be described as a combination of signal and noise $d_i =
S_i + \eta_i$. This noise is present due to the detection of a finite number of
photons by the microscope sensor, noise in the electronics, etc. and can be well
described for our system by uncorrelated $\langle \eta_i\eta_j\rangle = 2\sigma^2\delta_{ij}$,
Gaussian noise $\eta_i \sim \mathcal{N}(0, \sigma)$ (see
Section~\ref{sec:generative_model}).

In a Bayesian framework, the likelihood that an individual pixel is correctly
described by our model is given by the Gaussian likelihood,
\begin{equation}
\mathcal{L}(M_i \given d_i) = \frac{1}{\sqrt{2\pi \sigma_i^2}} e^{-(M_i - d_i) / (2\sigma_i^2)}
\end{equation}
For uncorrelated pixel noise, the entire likelihood of
the model given the image is given by the product over all pixels,
$\mathcal{L}(\vec{M} \given \vec{d}) = \prod_i \mathcal{L}(M_i \given d_i)$.
We are ultimately interested in the probability of the underlying parameters given
the image we record. According to Bayes' theorem, we can write this as
\begin{align*}
P(\vec{\theta} \given \vec{d}) &\propto P(\vec{d} \given \vec{\theta}) P(\vec{\theta}) \\
&\propto \mathcal{L}(\vec{M}(\vec{\theta}) \given \vec{d}) P(\vec{\theta})
\end{align*}
where $P(\vec{\theta})$ are priors that allow us to incorporate extra
information about the parameters $\vec{\theta}$. These priors can be as simple
as the fact that the particle radius is positive definite or that a group of
images share similar PSFs. For example, an overlap prior
$P_{\rm{overlap}}(\vec{x}_i, \vec{x}_j, a_i, a_j) = \Heaviside(a_i+a_j -
|\vec{x}_i - \vec{x}_j|)$, where $\Heaviside$ is the Heaviside step function,
can be used to impose the physical constraint that particles cannot overlap.
However, we found that the overlap prior only becomes relevant when the free
volume of a particle is small compared to the average sampling error volume
(when a particle is caged by $\sim1~\nm$ on all sides) and so we ignore it most
of the time.

We primarily work with the log-likelihood function $\log\mathcal{L}$ because
the number of pixels in the image can be very large, on the order $10^7$. For Gaussian noise, the
log-likelihood is precisely
the square of the $L_2$ norm between the model and the data. Therefore, we are
able to maximize this log-likelihood using a variety of standard routines
including linear least squares and a variety of Monte-Carlo sampling
techniques. After optimizing, we use the covariance $J^{T}J$ to determine errors
in the parameters or standard Monte-Carlo algorithms to sample from the
posterior probability distribution to extract full distributions of the model
parameters. In this way, any quantity of interest that is a function of particle
distribution can be calculated using Monte-Carlo integration by
\begin{align*}
\langle \mathcal{O}(\vec{\theta}) \rangle &= \int \mathcal{O}(\vec{\theta}) P(\vec{\theta} \given \vec{d})\diff\vec{\theta} \\
&= \frac{1}{N} \sum_i^N \mathcal{O}(\vec{\theta}^i)
\end{align*}

Here, $\vec{\theta}^i$ is a parameter vector sampled fairly from the
posterior probability distribution and $\mathcal{O}(\vec{\theta}^i)$ is an
observable such as the pair correlation function, packing fraction, or mean
squared displacement. Calculating higher-order moments provides estimated errors and error correlations on these observables. This is one of the more powerful aspects of
this method -- one can generate a probability distribution for each parameter
and directly apply these distributions to any observable that can be inferred
from the parameters.

Given this Bayesian framework, the main idea of this work is to create a full
generative model for confocal images of spherical particles and provide
algorithmic insights in order to implement the model on commodity computer
hardware.

\section{Generative model} \label{sec:generative_model}

Most of the difficulty in our method lies in creating a generative model that
accurately reproduces each pixel in an experimental image using the
fewest number of parameters possible. Our model is a physical description of
how light interacts with both the sample and the microscope optics to create the
distribution of light intensity that is measured by the microscope sensor and
rendered as an image on the computer.  In this section, we describe the model
which we use to generate images similar to those acquired by line-scanning
confocal microscopy of spherical particles suspended in a fluorescent fluid.

Our generative model aims to be an accurate physical description of the microscope
imaging; it is not a heuristic. Creating this model requires a detailed understanding of image formation of colloidal spheres in a confocal microscope. In the simplest view, our samples consist of a continuous distribution of dye distributed throughout the image. If the fluid is dyed (as for the images in this work), due to diffusion the dye is uniformly distributed through the fluid. The fluid-free regions, such as those occupied by the particles, are perfectly dye-free. 
The sample is illuminated with a laser focused through an objective lens. This
focused laser excites the fluorescent dye only in the immediate vicinity of the
lens's focus. An objective lens captures the dye's emitted light, focusing it
through a pinhole to further reject out-of-focus light. The collected light
passes through a long-pass or band-pass filter, which eliminates spurious reflected laser light before collection by a detector. This process produces an image of the sample at the focal point of the lens. Finally, rastering this focal region over the sample produces a three-dimensional image of the sample.

However, the actual image formation is more complex than the simple view
outlined above. Excessive laser illumination can cause the dye to photobleach.
Due to dirt and disorder in the optical train, the sample is not illuminated
uniformly. Diffraction prevents the laser light from being focused to a
perfect point and prevents the objective lens and pinhole from collecting
light from a single point in the sample. Aberrations
are present if the sample's refractive index is not matched to the
design of the objective lens, broadening the diffractive blur deeper into the
sample. Both the illuminating and fluorescing light can scatter off refractive
index heterogeneities in the sample due to the particles.

Some of these complications can be eliminated by careful sample preparation.
In practice, we eliminate photobleaching by using an excessive amount of dye
in our samples and illuminating with a weak laser light. We eliminate scattering by
matching the refractive index of the particles to the suspending fluid -- it is fairly easy to match the refractive indices to a few parts in $10^3$.
Since the scattering is quadratic in the index mismatch, the effect
of turbidity due to multiple-scattering is very weak in our samples. However, the rest of these complications must be accurately described by the generative model.

Based on this physical setup, we can describe the confocal images through three main generative model components:
\begin{itemize} 

\item \emph{Platonic image} $\Pi(\vec{x})$ -- the physical shape of the
dye distribution in the sample (unmodified by perception of light).

\item \emph{Illumination field} $I(\vec{x})$ -- the light
intensity as a function of position, including both laser intensity
variation from disorder in the optics and intensity
attenuation into the sample.

\item \emph{Point spread function} $P(\vec{x}; \vec{x}^{\prime})$ -- the image of a
point particle due to diffraction of light, including effects from index mismatch and
finite pinhole diameter.
\end{itemize}
plus three minor additional fit model components:
\begin{itemize}
\item \emph{Image Background} $c$, $B(\vec{x})$ -- the overall exposure of the image $c$ and the background values corresponding to a blank image without dye, $B$.

\item \emph{Rastering Step Size} $\zscale$ -- the displacement distance of the lens as it
rasters along the optical axis.

\item \emph{Sensor noise} $\sigma$ -- the noise due to shot noise from finite light intensity reaching
the sensor or electronic noise at the sensor.

\end{itemize}

These components are combined to form the image through convolution
\begin{equation} \label{eq:model}
\mathcal{M}(\vec{x}) = B(\vec{x}) + \int \diff^3x^{\prime}\,\,[I(\vec{x^{\prime}})(1-\Pi(\vec{x^{\prime}})) + c \Pi(\vec{x^{\prime}})] P(\vec{x}-\vec{x}^{\prime}; \vec{x})
\end{equation}
which is sampled at discrete pixel locations to give the final image $M_i =
\mathcal{M}(\vec{x}_i)$.

Here, we describe each part of our model in detail along with our explanations
and motivations behind any simplifications. In subsequent sections we
will also discuss other aspects of image formation which may result in other
model choices and why we omit them from the final form of the model.

\subsection{Platonic image}

The Platonic image must accurately represent the continuous distribution of fluorescent dye in the sample on the finite, pixelated image domain. The colloidal sample consists of a collection of spherical particles embedded in the solvent, with either only the particles or only the solvent dyed. Our Platonic image should then consist of the union of images of individual spherical particles, with their corresponding radii and positions. Thus, if we have a method to accurately represent one colloidal sphere, we can easily construct the Platonic image in our generative model.

A na{\"i}ve way to generate the Platonic image of one sphere would be simply to sample the dye distributions at the different pixel locations, with each pixel being either $0$ (if it is outside the sphere) or $1$ (if it is inside the sphere) with no aliasing. This method will not work, since a pixel value in the Platonic image can only change when a sphere's position or radii has shifted by one pixel. This method of Platonic image formation would produce a generative model that does not adequately distinguish between particle locations separated by less than 1 pixel or $100\nm$! Simply multiplying the resolution and corresponding coarse-graining of the boolean cut by a factor of $N$ in each dimension increases the resolution of this method to $1/N$ pixels. However, calculating these high resolution platonic spheres is computationally expensive, requiring $10^9$ operations to draw spheres capable of determining positions within $0.01\px$. 

To find the correct representation of a Platonic sphere, we examine the mechanism of image formation in Eq.~\ref{eq:model}. The final image results from a convolution of the Platonic image with the point-spread function $P(\vec{x}-\vec{x'}; \vec{x})$. Thus, we need a representation of a sphere that will produce the correct image after being convolved with the point-spread function. To do this, we recall that a convolution is a multiplication in Fourier space. However, creating the image of the sphere in Fourier space is problematic since there will be undesirable ringing in the Platonic image due to the truncation from the finite number of pixels (\textit{i.e.}~Gibbs phenomenon). Moreover, each update of one particle requires updating all the pixels in the image, which is exceedingly slow for large images.

Instead, we look for a functional form in real space that approximates the numerically-exact truncated Fourier series, where the truncation arises due to a finite number of pixels. For a sphere with radius $a$ at position $\vec{p}$, this truncated Fourier series is given by $\tilde{\Pi}(\vec{q}; \vec{p}, a) = 4\pi a^3 (j_1(q) / q) e^{i\vec{q}\cdot\vec{p}}$, where $\vec{q}$ is sampled only at frequencies in the image. We can view the truncation operation as a multiplication in Fourier space by a boxcar $\Heaviside(1-|q_x|)\Heaviside(1-|q_y|)\Heaviside(1-|q_z|)$, where $\vec{q}$ is the variable inverse to position, measured in $\px^{-1}$. By the convolution theorem, this truncation corresponds to a convolution in real space with $\sinc(x)\sinc(y)\sinc(z)$, using the inverse Fourier transform of the boxcar as the sinc function. Thus, the numerically exact image of a sphere would be the analytical convolution of $\sinc(x)\sinc(y)\sinc(z)$ with a sphere of radius $a$ at position $p$, represented on a discrete grid. However, the convolution with the sinc function is analytically intractable. To circumvent this, we approximate the sinc function by a Gaussian. This gives a representation of the correctly-aliased Platonic image $\Pi(\vec{x;a})$ of a sphere of radius $a$ as
\begin{equation} \label{eq:PI_def_integral}
\Pi(\vec{x}) = \mathcal{S}(\vec{x}) \ast \left[ \left( 2\pi \sigma_x^2 \sigma_y^2 \sigma_z^2 \right)^{-1/2} e^{-x^2/2\sigma_x^2}e^{-y^2/2\sigma_y^2}e^{-z^2/2\sigma_z^2} \right]
\end{equation}
where $\mathcal{S}(\vec{x}; \vec{p}, a) = H(|\vec{x}-\vec{p}| - a)$ where $H(x)$ is the Heaviside step function, which is either $0$ or $1$ depending on whether $|\vec{x} -\vec{p}| > a$ or $<a$, and $\ast$ denotes convolution. The Gaussian widths $\sigma$ should be approximately $1\px$; however, if the ratio of the $z$ pixel size to the $xy$ pixel size $\zscale \ne 1$, then $\sigma_z$ will not be the same as $\sigma_x$ and $\sigma_y$.

While Eq.~\ref{eq:PI_def_integral} does not generally admit a simple solution, there is a closed-form functional form for the symmetric case $\sigma_x = \sigma_y = \sigma_z$. In the symmetric case ($\zscale = 1$) Eq.~\ref{eq:PI_def_integral} takes the form
\begin{equation} \label{eq:PI_exact_gauss}
\Pi(\vec{x}) = \frac 1 2 \left[ \erf\left( \frac {a-r}{\sigma\sqrt{2}}\right) + \erf\left(\frac{a+r}{\sigma \sqrt{2}}\right) \right] - \frac 1 {\sqrt{2\pi}} \frac \sigma r \left[ e^{-(r-a)^2/2\sigma^2} - e^{-(r+a)^2/2\sigma^2} \right]
\end{equation}
where $r$ is the distance from the particle's center. The first bracketed group of terms corresponds to treating the sphere as a flat surface, and the second bracketed group corresponds to the effects the sphere's curvature on the integral. In each sub-grouping, the first term that depends on $r-a$ reflects the contribution due to the particle's nearer edge, and the second term that depends on $r+a$ reflects the contribution due to the particle's farther edge. We then fit $\sigma$ in Eq.~\ref{eq:PI_exact_gauss} to best match the exact Fourier space image of a sphere, giving a value $\sigma \approx 0.276$.

\begin{figure*} [tp]
\includegraphics[width=1.0 \textwidth]{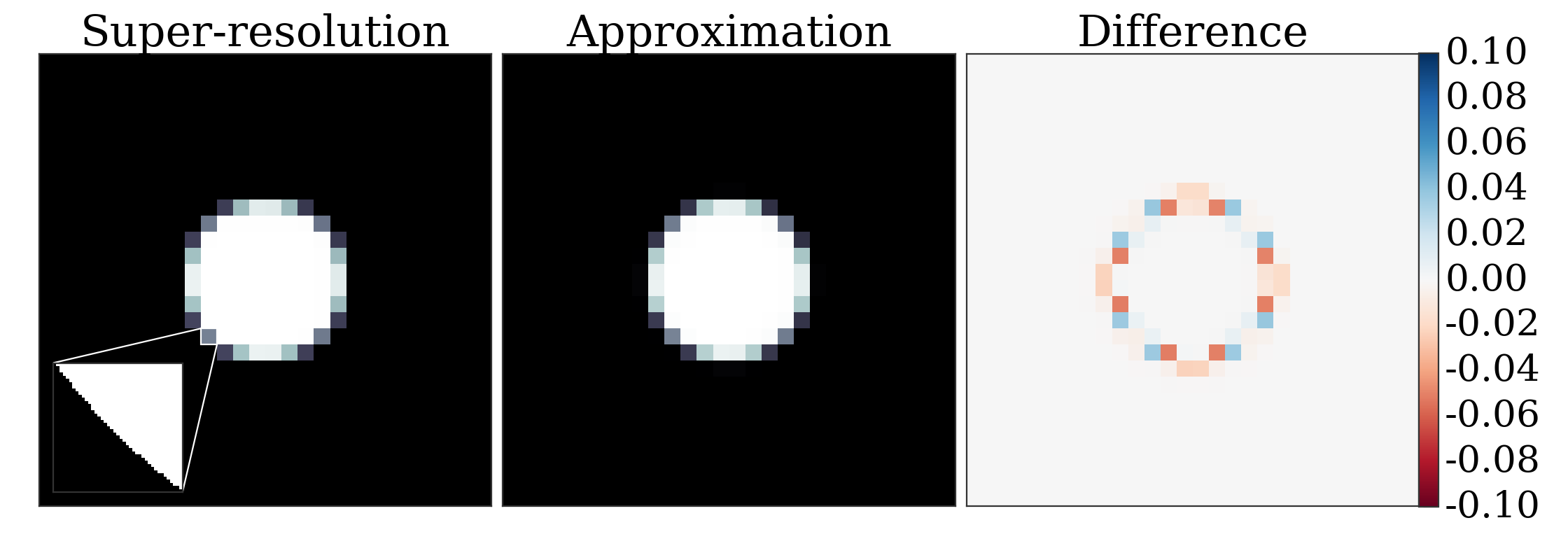}

\bcaption{Platonic sphere generation}{. A comparison of our approximate platonic sphere generation method to a sphere created by performing a boolean cut $\Pi(\vec{x}) = \int_{\rm{pixel}} \diff\vec{x}^{\prime} H(|\vec{x}-\vec{x}^{\prime} - \vec{p}| - a)$ on a lattice $100\times$ higher in resolution in each dimension compared to the final image. On the left we show the super resolution sphere with fractional volume error $\delta V/V = 10^{-6}$ and an inset displaying the jagged edges caused by discrete jumps in distance. This is in contrast to the iterative approximate platonic sphere with volume error $\delta V/V = 10^{-16}$ drawn at an effective radius with change $\delta a/a = 2\times 10^{-4}$. The differences between individual pixels along the center of the sphere (right panel) show a high frequency structure with a maximal relative value $0.08$. These high frequency features are dramatically reduced later in the image formation process through the convolution with the point spread function.}

\label{fig:platonic_spheres}
\end{figure*}

Although Eq.~\ref{eq:PI_def_integral} does not admit a simple solution for $\zscale \ne 1$, we can use the exact form for $\zscale = 1$ to construct an approximate solution. Since both $\erf(x)$ and $e^{-x^2}$ approach their asymptotic values extremely rapidly, and since at the best fit $\sigma \approx 0.276$
$(a+r)/\sigma \gg 1$ for even moderately small radii, the terms $\erf( (a+r)/\sigma \sqrt{2}) \approx 0.5$ and $\exp( -(r+a)^2/2\sigma^2) \approx 0$ to an excellent accuracy. We then write the position vector in terms of its direction $\hat{x}$ and a vector $\vec{\delta x}$ as $\vec{x} \equiv a \hat{x} + \vec{\delta x}$, and replace $(a-r)/\sigma$ in Equation~(\ref{eq:PI_exact_gauss}) by $\sqrt{ (\delta x / \sigma_x)^2 + (\delta y / \sigma_y)^2 + (\delta z / \sigma_z)^2 }$. Note that this approximation is exact in the limit of infinite sphere radii. Empirically, we find that this approximation works quite well, giving differences in the Platonic image of a few percent from a numerical solution to Eq.~\ref{eq:PI_def_integral} as well as high resolution boolean cut real-space spheres (see Fig.~\ref{fig:platonic_spheres}).

While this implementation of the Platonic image correctly captures most of the effects of finite-pixel size, there are still some minor details that need to be fixed to give unbiased images. By construction, Eq.~(\ref{eq:PI_exact_gauss}) conserves volume -- its integral over all space is $4/3 \pi a^3$ since the Gaussian kernel is normalized. However, when $\Pi(\vec{x})$ is sampled on a pixelated grid, its sum is not exactly $4/3 \pi a^3$ but is slightly different, depending on the position of the particle's center relative to a voxel's center. The slight change in volume is important for two reasons. First, the convolution with the PSF in our image generation (see next subsection) suppresses high-frequency portions of the image, but it does not affect the $\vec{q}=\vec{0}$ component, \textit{i.e.} the image sum or the particle volume. Since we aim to create a Platonic image that accurately represents the final image, we need the $\vec{q}=\vec{0}$ component of the Platonic image to be correct. Secondly, as discussed in section~\ref{sec:model_selection} the real microscope image is actually an integral over a finite pixel area. As such, the image recorded on the detector preserves the particle's volume or the $\vec{q}=\vec{0}$ component of the image. To circumvent this issue of incorrect particle volume, instead of drawing the particle at its actual radius we draw it with a slightly different radius that preserves the particle's volume, which we accomplish with an iterative scheme. The results of this iterative scheme are shown in Fig.~\ref{fig:platonic_spheres} along with the errors it introduces. Incidentally, the effects of image pixelation on image moments higher than $\langle 1 \rangle$, \textit{e.g.} $\langle \vec{x} \rangle$ and its effects on the particle positions, are much smaller than the noise floor in our data at a moderate SNR (see section~\ref{sec:model_selection}).

The representation in equation~\ref{eq:PI_exact_gauss} is the best method for forming Platonic spheres on a pixelated grid that we have found. However, there are other, simpler methods which work almost as well as the Platonic sphere. Aside from the important curvature term, equation~\ref{eq:PI_exact_gauss} is basically an $\erf()$ interpolation between particle and void at the particle's edge. Other interpolation schemes can provide similar results. For instance, the spheres could be constructed by ignoring the curvature term and replacing the $\erf$ with a logistic $1/(1+\exp((r-a)/\alpha))$, a linear interpolation between particle and void at the pixel edge, or a cubic interpolation at the pixel edge. We have also implemented these methods for generating Platonic images of spheres, fitting the parameters to match the exact Fourier representation. For the logistic we fit $\alpha$, for the linear interpolation we fit the slope, and for the cubic we fit one parameter and constrain the other two such that the Platonic image and its derivative are continuous. While all of these methods are functional, they are not significantly faster than the exact Gaussian approximation in equation~\ref{eq:PI_exact_gauss} and result in slightly worse featuring errors (see table~\ref{table:model-complexity}). As a result, we use the exact Gaussian approximation, but include these other options in our package for ease of use with more complicated shapes where the integral in equation~\ref{eq:PI_def_integral} might not be analytically tractable.

The Platonic image needs to represent accurately all objects in the image, not just the spheres. In particular, when the solvent is dyed, the image usually contains a dark coverslip or its shadow from the point-spread function. We model this dark coverslip as a slab occupying a half-space. The slab is characterized by a $z$-position and by a unit normal $\hat{n}$ denoting the perpendicular to the plane. To capture accurately sub-pixel displacements of the slab, we use the image of a slab convolved with a Gaussian as above for a sphere; for the slab this gives a simple error (erf) function.

\subsection{Illumination field}

In order to illuminate the sample, confocal microscopes scan a laser over the
field of view using several distinct patterns including point, line, and disc
scanning.  This illumination laser travels through the optics train and
interacts with fluorescent dye in the suspension causing it to emit light in a
second wavelength which is then detected.  The intensity of this illumination
pattern depends on the aberrations in the optics as well as dirt in the optical
train which creates systematic fluctuations in illumination across the field of
view. Accounting for these variations is important as they can account for most of the intensity variation in an image.  In the case of our line scanning confocal microscope, these patterns manifest themselves as stripe patterns perpendicular to the scan direction, as the line-scan drags dirt across the field of view, overlaid on aberrations and optical misalignments which cause the corners of the image to dim.

\begin{figure*} [ht]
\includegraphics[width=1.0 \textwidth]{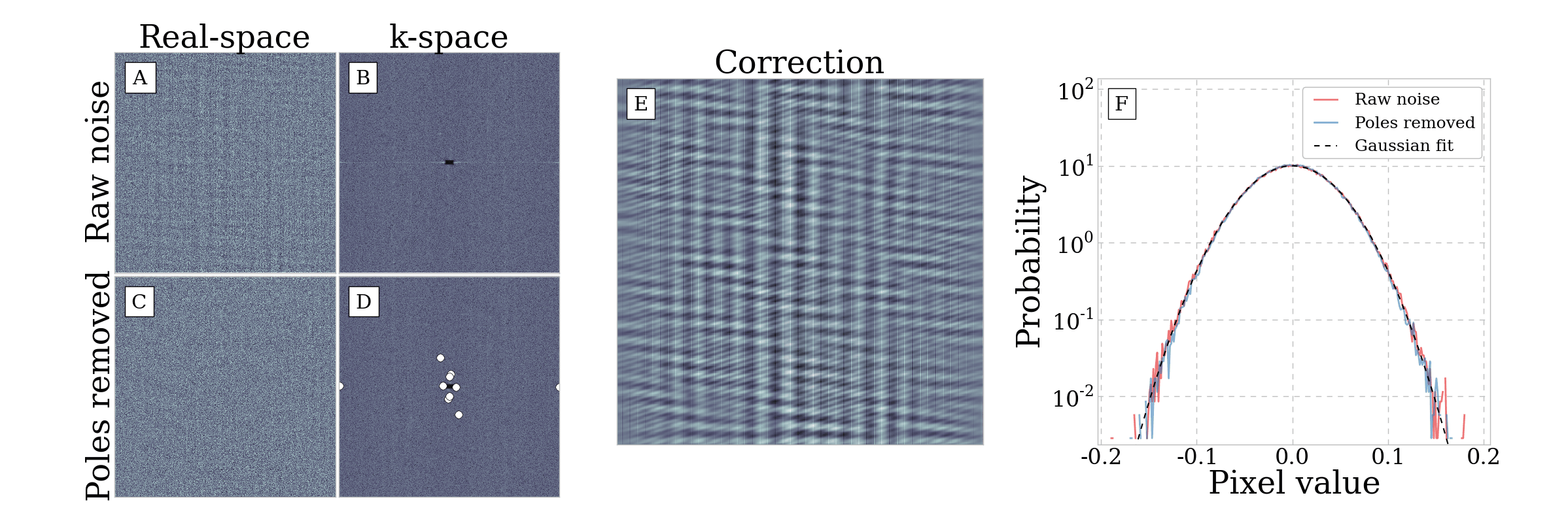}

\bcaption{Illumination field residuals}{. A blank confocal image and its fit to the Barnes ILM in equation~\ref{eq:barnes_ilm} over varying number of coefficients. Fitting the illumination with a low-order ILM of $(3,3)$ Barnes points removes the large fluctuations over the image but clearly shows stripes in the image. The notation $(n_0, n_1, n_2,...)$ corresponds to a Barnes ILM with $n_0$ coefficients in the expansion for $P_0(y)$, $n_1$ coefficients for $P_1(y)$, etc. Increasing the number of points to $(7, 7, 5, 5, 5)$ or $(14,9, 7, 5, 5, 5)$ removes the overall modulation in $y$ but leaves clear stripes in the image. Only at high orders of $(50,30,20,12,12,12,12)$ or $(200,120,80,50,30,30,30,30,30,30,30)$ do these stripes disappear. The residuals shown in the figure are all at the same scale and are averaged over the image $z$ for clarity.
}

\label{fig:ilm_residuals}
\end{figure*}

Confocal microscopes image by rastering in $z$, illuminating each $xy$ plane separately. Ideally, the microscope illuminates each plane identically. In practice, aberrations due to refractive index mismatches cause a dimming of the illumination with depth into the sample~\cite{Hell1993}. Since this overall dimming only depends on the depth $z$ from the interface and not on the $xy$ position in the sample, it is natural to describe the illumination field as a product of an $xy$ illumination and a $z$ modulation:
\begin{equation}
I(\vec{x}) = I_{xy}(x,y) \times I_z(z) \quad .
\end{equation}
Empirically we find that illumination fields of this form can accurately describe our real confocal images, without incorporating any coupling between $xy$ and $z$.

We describe each of the separate functions $I_{xy}$ and $I_z$ by a series of basis functions. Since the modulation in $z$ is fairly smooth~\cite{Hell1993}, we describe $I_z(z)$ by a polynomial $P_z(z)$ of moderate order $\approx$ 7-11 for 50-70 $z$-slices; typically we use a Legendre polynomial as the orthogonality accelerates the fitting process. The in-plane illumination of a confocal is determined by its method of creating images. Our confocal is a line-scanning confocal microscope, which operates by imaging a line illumination parallel to the $x$ axis and simultaneously collecting the line's fluorescent image. This line is then scanned across the image in $y$. As a result of this scanning, any dirt in the optics is dragged across the field of view, creating the illumination with stripes along the $x$-direction visible in Fig.~\ref{fig:ilm_residuals}. To model these stripes, we treat the variation along $x$ and $y$ differently. We write the $xy$ illumination field as
\begin{equation}
I_{xy}(x,y) = \sum_k B(x; \vec{c}_k) \times P_k(y) \quad ,
\label{eq:barnes_poly_xy}
\end{equation}
where $B_k(x;\vec{c}_k)$ is a Barnes interpolant in $x$ and $P_k(y)$ a Legendre polynomial in $y$. Barnes interpolation is a method of interpolating between unstructured data using a given weight kernel~\cite{barnes1964technique}, similar to inverse distance weighting, using a truncated Gaussian kernel to allow for strictly local updates to the high frequency illumination structure. We use an interpolant with equally spaced anchor points in $x$ throughout the (padded, see section~\ref{sec:gm_psf}) image. The $k^\textrm{th}$ Barnes interpolant has a large number of free parameters, described by the vector $\vec{c}_k$; the size of $\vec{c}_k$ is equal to the number of anchoring points in the Barnes. To account for the fine stripes in the image, we use a large number of points for the Barnes associated with low-order polynomials, and decrease the number of points for higher-order polynomials. For a typical image of size $(z,y,x)=(50,256,512)$ pixels, we use coefficient vectors of length $(\vec{c}_0, \vec{c}_1, \vec{c}_2, \vec{c}_3, \vec{c}_4, \vec{c}_5, \vec{c}_6, \vec{c}_7, \vec{c}_8, \vec{c}_9, \vec{c}_{10}) \approx (200,120,80,50,30,30,30,30,30,30,30)$. While this is a large number of coefficients, there are orders of magnitude fewer coefficients than pixels in the image. As a result, all of the ILM parameters are highly constrained (on the order of a few parts in $10^5$, varying wildly with the parameter), and we do not overfit the image.

Putting this all together, we use an ILM given by:
\begin{equation}
\left[
\sum_k B_k(x; \vec{c}_k) P_k(y)
\right] \left[\sum_j d_j P_j(z)\right] \quad .
\label{eq:barnes_ilm}
\end{equation}
This ILM accurately describes measured confocal illuminations, as determined both from blank images and from images with colloidal particles in them. While the Barnes structure of this ILM is optimized for line-scanning microscopes, it can easily be changed. For ease of use for different microscopes or imaging modalities we have implemented various ILMs consisting of simple Legendre polynomial series, as functions $P_{xy}(x,y) \times P_z(z)$, $P_{xy}(x,y) + P_z(z)$, and as $P_{xyz}(x,y,z)$. Other illumination structures -- such as a radially or azimuthally striped ILM for spinning-disk confocals -- could also easily be incorporated into PERI's framework.

How well do these functional forms fit to experimental data for a line-scanning
confocal microscope?  We acquire blank images of a water-glycerol mixture as a
function of depth and fit this data with Barnes illuminations of the form Eq.~\ref{eq:barnes_ilm}.  As a function of the number of Barnes points in $x$ and the polynomial degree in $y$, we look at the magnitude and patterns of the residuals. In Fig.~\ref{fig:ilm_residuals}, we see large scale structure in the ILM residuals, suggesting that high-order polynomials and Barnes interpolants with a large number of points are necessary. Fitting out the low-order background reveals the find stripes in $x$ emerge due to the line-scan nature of our machine. Finally, at  higher orders of interpolants and polynomials we are able to adequately capture all illumination variation independent of depth into the sample.

\begin{figure*} [htp]
\includegraphics[width=0.75 \textwidth]{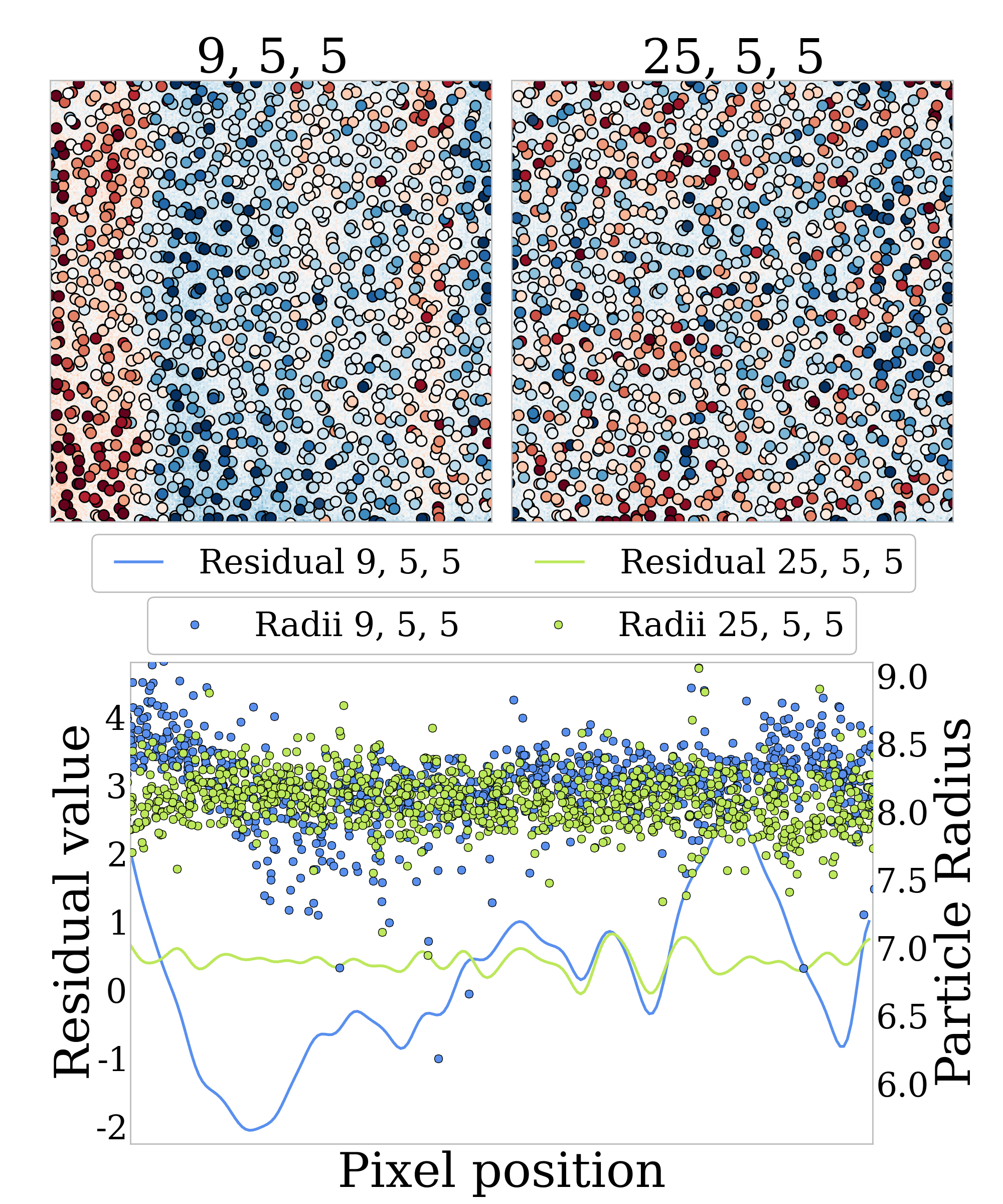}

\bcaption{ILM generated biases}{. Using an incorrect illumination field results in significant biases. The upper left panel shows an image featured with a $(9,5,5)$ order polynomial in $(x,y,z)$. In the foreground are the featured particle radii, color-coded according to their difference from the mean. In the background is the residuals of the featured image. Clear stripes are visible in both the featured radii and the residuals. The particles are systematically much larger on the left side of the image, before decreasing in size in the middle and increasing again in a small stripe on the image's right side. In contrast, when the image is featured with a higher-order $(25,5,5)$ degree polynomial, shown in the upper right, these systematic residuals disappear. The bottom panel shows the particle radii and image residuals for the two illumination fields as a function of the image $x$ direction.
}

\label{fig:ilm_smile}
\end{figure*}

Fitting the ILM correctly is essential for finding the correct particle
positions and radii. Fig.~\ref{fig:ilm_smile} demonstrates the effect of
featuring a real confocal image with an illumination field of insufficient
order. In the left panel is an image featured with a high-degree polynomial
illumination of 9\textsuperscript{th} order in the $x$-direction and of
5\textsuperscript{th} order in the $y$- and $z$- directions. While these
polynomials are high-order, they are not high enough to capture all of the
structure in the light illumination. There is a clear bias in the featured
radii, with particle radii being systematically larger on the edge of the image
and smaller in the middle. These biases arise from large stripes in the
confocal illumination due to the line-scanning nature of our confocal. Using a
higher-order 25\textsuperscript{th} degree polynomial in the $x$-direction
(upper right panel) eliminates the effect of these stripes, as visible in the
featured particle radii plotted as a function of $x$ in the bottom panel. Note
that the particle radii may be biased by as much as $1\px$ or $100\nm$ due to
effects of the spatially varying illumination field.

\subsection{Point spread function} \label{sec:gm_psf}

Due to diffraction, the illuminating laser light focused from the microscope's lens and the detected fluorescent light collected from the sample are not focused to a single point. Instead, the light is focused to finite-sized diffraction-limited blur. To reconstruct an image correctly we need to account for the effects of diffraction in image formation.

A confocal microscope first illuminates the sample with light focused through the microscope lens. The lens then collects the light emitted from fluorophores distributed in the sample. As a result, the final image of a point source on the detector results from two separate terms: an illumination point-spread function $P_\textrm{ilm}$ that describes the focusing of the incoming laser light, and a detection point spread function $P_\textrm{det}$ that describes the focused fluorescent light collected from the emitted fluorophores. Since a fluorophore is only imaged if it is both excited by the laser illumination and detected by the camera, the resulting point-spread function for a confocal with an infinitesimal pinhole is the product of the illumination and detection point-spread functions: $P(\vec{x}) = P_\textrm{ilm}(\vec{x}) P_\textrm{det}(\vec{x})$. For a confocal with a finite-sized pinhole, this product becomes an convolution over the pinhole area. The two separate point-spread functions (PSFs) $P_\textrm{ilm}$ and $P_\textrm{det}$ can be calculated from solutions to Maxwell's equations in the lens train \cite{Hell1993, Visser1994, Zhang2007, Nasse2010}. The PSFs can be written as integrals over wavefronts of the propagating light.

An additional complication arises from the presence of an optical interface. Most microscope lenses are essentially ``perfect'' lenses, creating a perfect focus in the geometric optics limit. However, refraction through the optical interface destroys this perfect focus and creates an image with spherical aberration. In addition, the refracted rays shift the point of least confusion of the lens from its original geometric focus. For a confocal geometry, this spherical aberration and focal shift depend on the distance of the nominal focal point from the optical interface $\zint$.

\begin{figure*} [tp]
\includegraphics[width=1.0 \textwidth]{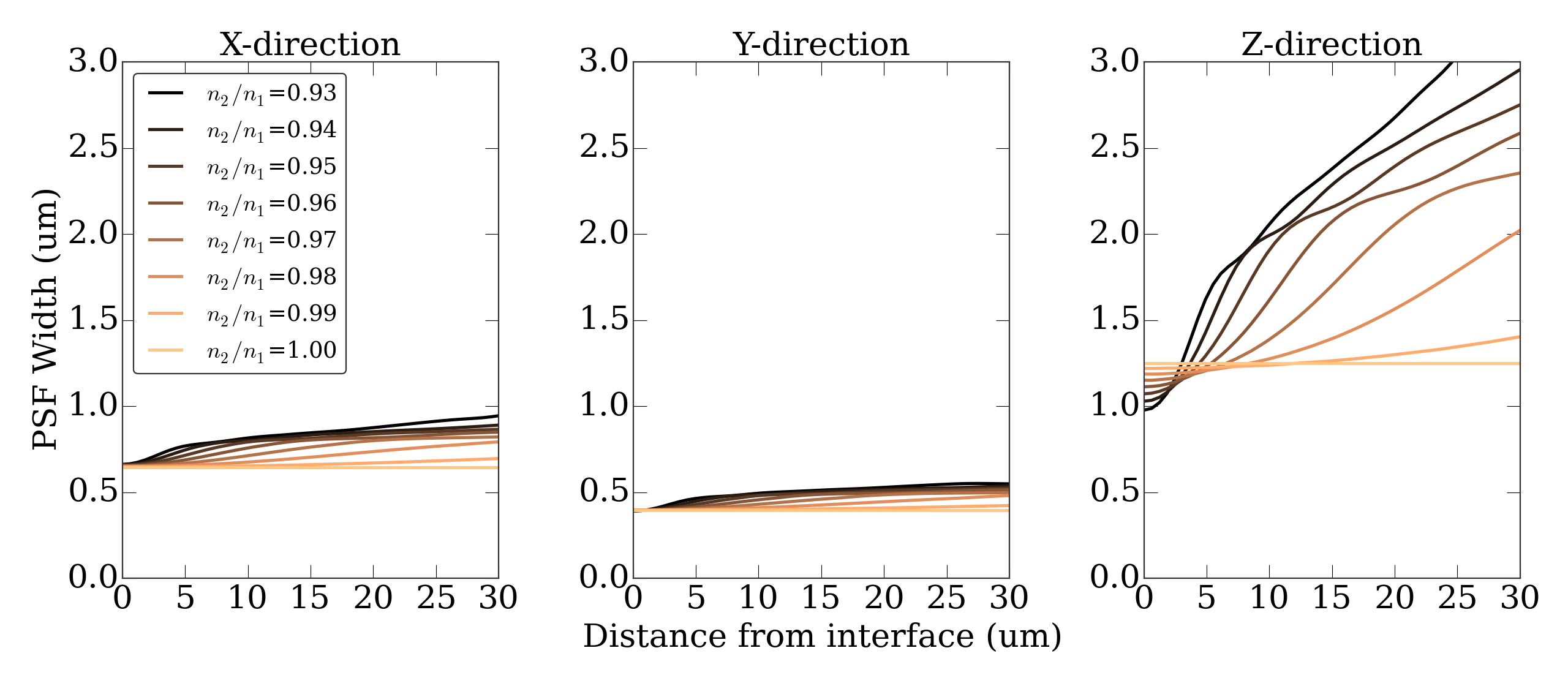}

\bcaption{PSF widths vs depth}{. The $x$ (left panel), $y$ (center panel), and $z$ (right panel) widths of the PSF as a function of distance from the interface, for various refractive index mismatches. The width of the point-spread function generally increases with depth and with index mismatch due to increased spherical aberrations. The width is broadest in the $z$ (axial) direction, and is narrower in the $y$ direction than along the $x$ direction of the line illumination.
}

\label{fig:gaussian_fit_exact_psf}
\end{figure*}

All of these effects have been calculated in detail by many previous researchers \cite{Hell1993, Visser1994, Zhang2007, Nasse2010}. The PSFs depend on several parameters: the wave vectors of the incoming and outgoing light $k_{\rm{in}}$ and $k_{\rm{out}}$, the ratio of the indices of refraction $n_{\rm{sample}} / n_{\rm{lens}}$ of the sample and the optical train design, the numerical aperture of the lens or its acceptance angle $\alpha$, and the distance focused into the sample $\zint$. For completeness, we repeat the key results here. In polar coordinates, the illumination PSF $P_\textrm{ilm}(\rho, \phi, z)$ for illuminating light with wave vector $k_\textrm{in}$  traveling through a lens focused to a depth $\zint$ from the interface is \cite{Hell1993}
\begin{equation} \label{eq:PSFcalc}
\begin{aligned}
P_\textrm{ilm}(\vec{x}) &= \left| K_1 \right|^2 + \left| K_2 \right|^2 + \frac 1 2 \left| K_3 \right|^2 + \cos 2 \phi \left[ K_1 K_2^* + K_2 K_1^* + \frac 1 2 \left| K_3 \right|^2 \right] \quad \textrm{, where} \\
\left( \begin{array}{c} K_1 \\ K_2 \\ K_3 \end{array} \right) &= \int_0^\alpha \sqrt{\cos \theta' } \sin \theta' e^{-ik_\textrm{in} f(z,\theta')} \left( \begin{array}{c} \frac 1 2 (\tau_s(\theta') + \tau_p(\theta') \cos \theta_2) J_0(k_\textrm{in} \rho \sin \theta') \\ \frac 1 2 (\tau_s(\theta') - \tau_p(\theta') \cos \theta_2) J_2(k_\textrm{in} \rho \sin \theta') \\ J_1(k_\textrm{in} \rho \sin \theta') \tau_p (\theta') \frac {n_1}{n_2} \sin \theta' \end{array} \right) \, d\theta'
\\
f(\theta) &= \zint \cos{\theta} - \frac{n_2}{n_1} (\zint - z)\sqrt{1 - \left(\frac{n_1}{n_2}\right)^2\sin^2\theta}
\end{aligned}
\end{equation}

Here $\tau_s(\theta')$ and $\tau_p(\theta)$ are the Fresnel reflectivity coefficients for $s$ and $p$ polarized light, $J_n$ is the Bessel function of order $n$, and $\theta_2$ is the angle of the refracted ray entering at an angle $\theta'$ ($n_2 \sin \theta_2 = n_1 \sin \theta'$). To derive this equation from equation~(12) in Ref.~\cite{hell1993aberrations}, we used the additional assumption that all distance scales in the image (including $\zint$) are small compared to the focal length of the lens. The corresponding detection PSF $P_\textrm{det}$ is identical to $P_\textrm{ilm}$ except for the removal of the $\sqrt{\cos\theta}$ and the replacement of $k_\textrm{in}$ by the wave vector of the fluorescent light $k_\textrm{out}$. For an infinitesimal pinhole, the complete PSF is the product of these two point spread functions:
\begin{equation}
P(\vec{x}; \zint) = P_{\rm{ilm}}(\vec{x}; \zint) P_{\rm{det}}(\vec{x}; \zint) \quad .
\label{eq:psf_prod}
\end{equation}

The expressions in equations~\ref{eq:PSFcalc}-\ref{eq:psf_prod} are for a perfect pinhole confocal, whereas our confocal is a line-scanning confocal. While there have been several works describing line-scanning confocals \cite{Wolleschensky2005, Dusch2007}, these authors have treated where the line is focused onto the sample by a cylindrical lens. In our confocal, however, an image of a line is focused onto the sample through the large-aperture objective lens. As such, the illumination PSF in equation~\ref{eq:psf_prod} is replaced by the integral of the detection PSF over a line in the $x$ direction.

We use this model for a line-scanning point spread function with aberrations as our model for our exact PSF, fitting the paramters that enter into equations~\ref{eq:PSFcalc}-\ref{eq:psf_prod}. These parameters are the acceptance angle $\alpha$ of the objective lens, the wavelength of the laser, the ratio of energies of the fluorescent light to the excitation light, the index mismatch $n_1/n_2$ of the sample to the optics, the position of the optical interface $\zint$, and the amount that the lens is moved as the scan is rastered in $z$. In principle, other details could be included -- polychromaticity and distribution of the fluorescent light, finite pinhole width of the illuminating line, etc. -- but we find that these parameters are both relatively unconstrained by the fit and have little impact on the other reconstructed parameters, such as particle positions and radii.

In addition, for initial featuring we occasionally use a Gaussian approximation to the PSF. Based on calculations of the exact PSF, $\approx 90\%$ of the function can be described by a Gaussian~\cite{Zhang2007}. We verified this for PSFs calculated from Eq.~\ref{eq:PSFcalc}, and found that although the presence of aberrations from the interface worsens the Gaussian approximation, generally a Gaussian accounts for $\approx 90\%$ of the PSF except for in the most aberrated cases (large index mismatch imaging deep into the sample). Our simplest approximation of the PSF is as an anisotropic Gaussian with different widths in $x,\,y,\,\textrm{and } z$, with the widths changing with distance from the interface.
We therefore parameterize the Gaussian widths as a function of depth,
\begin{equation}
P(\vec{x}; z) = \prod_i \frac{ e^{-x_i^2/2\sigma_i^2(z)} }{ \sqrt{2\pi} \sigma_i(z)}
\end{equation}
where each width $\sigma_i(z)$ is described by a polynomial in $z$, typically a second order Legendre polynomial.

\begin{figure*} [htp]
\includegraphics[width=1.0 \textwidth]{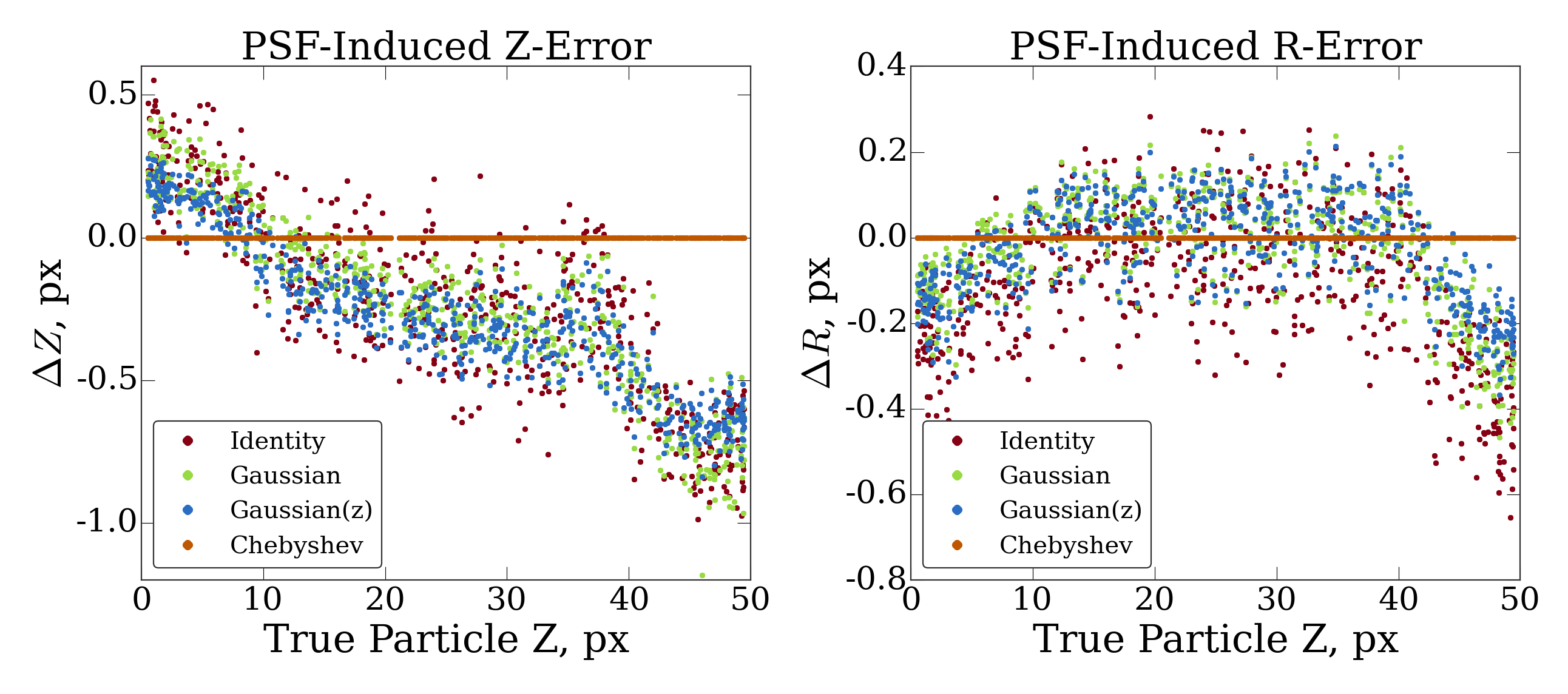}

\bcaption{PSF generated biases}{. Using an incorrect point-spread function results in significant biases, as PSF leakage affects neighboring particle fits. Moreover, since the PSF gets significantly broader with depth, using a spatially constant PSF, there are systematic biases with depth in both the $z$ positions (left panel) and a characteristic drift in the fitted radii errors with depth (right panel), as shown for the delta-function (identity), an $(x,y,z)$ anisotropic Gaussian, and a depth-varying Gaussian point-spread function. In contrast, using the correct Chebyshev PSF eliminates the errors in both the radii and $z$ positions (data points forming thin orange line).}

\label{fig:psf_smile}
\end{figure*}

Figure~\ref{fig:psf_smile} shows the effects of ignoring these details about the point-spread function on the extracted positions. We generate confocal images using a simulated, exact PSF with random distribution of particles up to a depth of $30\micron$. Featuring this data using a 3D anisotropic Gaussian, we find a strong depth-dependent bias in the featured $z$ position and radii measurements. Using a low order $z$-dependent Gaussian PSF decreases this bias only slightly. Interestingly however, ignoring the effects of diffraction completely and replacing the PSF with a Dirac delta-function does not cause significantly worse results than treating the PSF as a spatially-varying Gaussian. As shown by Fig.~\ref{fig:psf_smile}, an exact PSF is required to locate particle's positions and radii to within 20~nm (0.2~px). Therefore, we employ the full line-scan PSF calculation into our model.

The point-spread function defined in equations~\ref{eq:PSFcalc}-\ref{eq:psf_prod} decays extremely slowly with $z$ and somewhat slowly in $\rho$. To accurately capture these long-tails of the PSF in our generative model, we calculate the PSF on a very large grid for convolutions, corresponding to $\tightapprox~40 \tighttimes 25 \tighttimes 30 \px$ or $\tightapprox 6 \tighttimes 3 \tighttimes 4 \micron$ in extent, which is considerably larger than the size of the 5~px radii particles. The long tails of the PSF bring information about structure far outside the image into the image region. As such, our generative model is defined not only in regions corresponding to the interior microscope image but also in an exterior padded region, which is cropped out when comparing to the model. For completeness, we still define the ILM and Platonic image (including exterior particles) in the exterior padded region; however parameters confined to this exterior region of the image are relatively unconstrained. We make up for this loss in speed due to the increased size by doing an extremely accurate but approximate convolution based on Chebyshev interpolation, as described in a future paper.

\subsection{Background}

\begin{figure*} [tp]
\includegraphics[width=0.6 \textwidth]{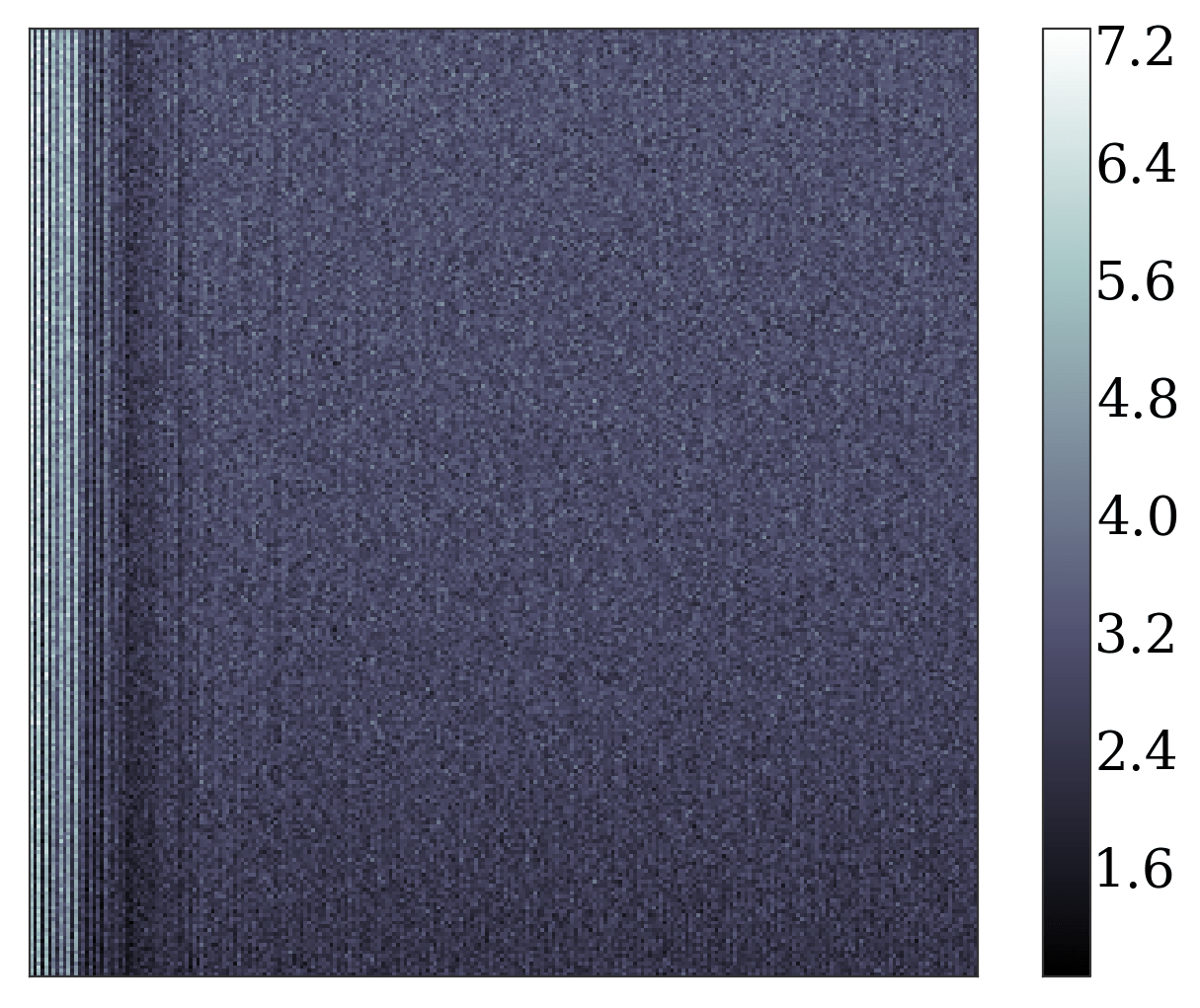}

\bcaption{Experimental background image}{.
The measured background from our line-scan confocal microscope captured by
adjusting the exposure to a full brightness image, removing the sample, and
capturing a set of images with no illumination including room lights. Note
that the range of values is from $1$ to $7$ out of a maximum $255$ given by
the $8$-bit resolution of the CCD. While only a variation of $3\%$, we have
seen in the illumination field section that this can create a bias that
significantly alters our inference as a function of the position in the
field of view. To remove this bias we fit the background field to a low
order polynomial and add it to our model image.
}

\label{fig:bkg_noise}
\end{figure*}

Due to background, the detector CCD pixels always read a non-zero value even
when there is no light incident on them. We incorporate this into our
generative model by fitting a nonzero background level to the images.  Ideally,
this background would be constant at every pixel location.  Empirically,
however, we find from blank images that this background varies with pixel
location in the detector (see Fig.\ref{fig:bkg_noise}). For our confocal
microscope, we find the background is slowly-varying in the optical plane,
perhaps due to different dwell times for different regions of the line scan and different sensitivies of different pixels; the background does not vary in $z$. As a result, the background is well-modeled by a
low-order polynomial in $x$ and $y$.

However, due to the long-tails of the PSF, the coverslip slab
affects the image in a much larger $z$ region than that of a typical particle.
Rather than dealing with this by using an even larger point-spread function, we use the calculated
point spread function to capture the effects of the PSF's moderate tails
on the particles and slab, and fit a polynomial in $z$ to capture the residual
slab correction. This residual correction is mathematically the same as a
background level in the detector. As a result, while the ``true'' background in
the image is $P(x,y)$, our model uses a background $P(x,y) + P_{slab}(z)$, as the coverslip is usually oriented along the $z$ direction.

\subsection{Sensor noise}

\begin{sidewaysfigure} [tp]
\includegraphics[height=0.34 \textwidth]{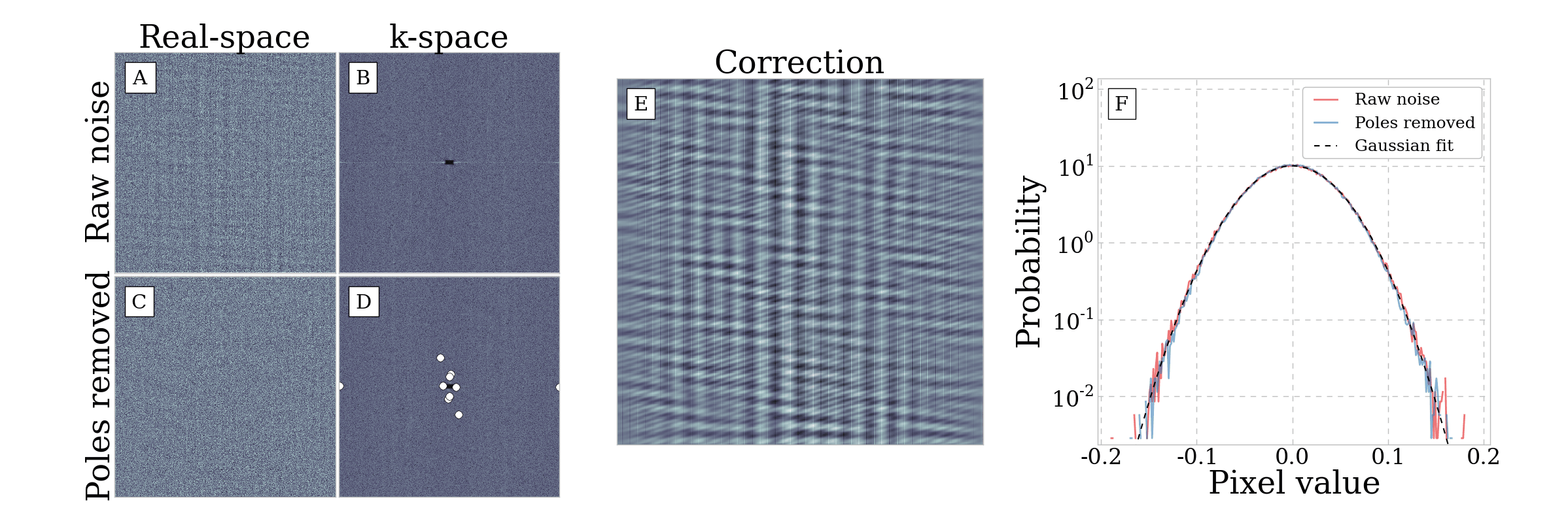}

\bcaption{Noise spectrum}{ (a) Real-space plot of residuals representing
the intrinsic noises generated in line-scanning confocal microscopy. This
noise spectrum was generated by subtracting the background from a blank
sample as in~Fig.\ref{fig:ilm_residuals}. Notice that while most of the
signal appears to be white noise, there is a systematic modulation along
the $x$ coordinate and high frequency features in the $y$ scan direction.
(b) Fourier power of the noise spectrum given in (a). The high frequency
modulation can now be seen as two small `poles' in the Fourier power
spectrum.  Note the dark box in the center of the spectrum is created by
subtracting the high order polynomial background from the blank image. In
(c) and (d) we present the real and Fourier space noise after removing
several discrete peaks in the Fourier intensity that represent correlated
noise sources.  The removed signal can be seen in (e) showing the stripes
created by the scanning nature of the confocal microscope.  In (f) we show
the histogram of residuals from (a) and (c).  In solid red we plot the data
and in dashed black lines we plot a Gaussian fit 
to the residuals with a
width $\sigma=0.0398$, showing that the noise spectrum is well approximated
by a Gaussian distribution after taking into account long wavelength
background features.
}

\label{fig:ilm_residuals_noise}
\end{sidewaysfigure}

The last feature of the generative model is our understanding of the
unrecoverable parts of the image: noise. To study the intrinsic noise spectrum
of the confocal microscope, we subtract the long wavelength behavior from the
blank image of Fig.~\ref{fig:ilm_residuals}.  After removing the background we
find that the noise appears white and is well approximated by a Gaussian
distribution (see Fig.~\ref{fig:ilm_residuals_noise}).  There are, however, some
highly localized non-Gaussian parts to the noise spectrum, arising due to the
specific nature of our confocal.
For instance, at high scan speeds slight intensity fluctuations in
the laser's power couple to the dwell time on each stripe of line-scanned pixels. This
produces periodic stripes across the image with a wavevector mostly parallel to the
scan direction, but with a random noisy phase.
How can we handle these sources of correlated noise and do they affect the quality of our
reconstruction?

In principle, these correlated noise sources can be represented in the Bayesian model
by introducing a full noise covariance matrix.  That is, instead of writing
that log-likelihood as the product of all pixel values, we can write
\begin{equation}
\log\mathcal{L}(\vec{M} \given \vec{d}) = -\frac{1}{2} (M_i-d_i) \,\Lambda^{-1}_{ij}\, (M_j-d_j)
\end{equation}
where $\Lambda^{-1}_{ij}$ is the covariance matrix between each pixel residual
in the entire image.  In our optimization, we would form a low dimensional
representation for this covariance matrix and allow it to vary until we
find a maximum.  In doing so, we would reconstruct the image and the
correlated noise simultaneously.  In practice, this introduces a large
computational overhead due to the need for a full image convolution during
each update as well as many new free parameters that need to be optimized.

Therefore, when desired we address the effect of correlated noise by working in reverse -- we identify the
several intense Fourier peaks in the confocal noise spectrum and remove them
from the raw data before the fitting process.  An example of this noise pole removal is given in
Fig.~\ref{fig:ilm_residuals_noise}. There, we can see that removing only 5
distinct poles (Fig.~\ref{fig:ilm_residuals_noise}(d)) removes almost all
visible correlated noise structure while changing the overall noise magnitude
by a negligible amount.  This small shift in estimated noise magnitude only
affects the estimate of the errors associated with parameters such as positions
and radii in a proportional way.  Since these errors are very small
and do not bias our inferred parameters, we often ignore the confocal's noise poles in our analysis
entirely.

\section{Model considerations} \label{sec:model_selection}

Here, we investigate several complexities of image formation in confocal
microscopes and systematically analyze whether or not it is necessary to
include them in our generative model.  In particular, we will first analyze how
much complexity we must introduce into the model elements listed in the
previous section, including the platonic image, illumination field, and point
spread function.  We will also look at elements of image formation which we
have not explicitly included in our model. First, confocal microscopes
build a 3D image by rastering in 1, 2, or 3 dimensions (see section~\ref{sec:generative_model}). There is
noise in this rastering procedure that affects the image formation process. Second,
The final image that comes from this scan is a cropped view of a much larger
sample; the edges of this cropped image are influenced by the excluded exterior
particles.  Third, while the actual distribution of light intensity is a continuous field, the detector only measures a pixelated representation of this field. Fourth, while the exposure is made by the camera, particles undergo diffusional motion, blurring their apparent location. In this section, we address each of these image formation complexities and their effects on the inferred parameters.

\begin{sidewaysfigure} [tp]
\includegraphics[height=0.40 \textwidth]{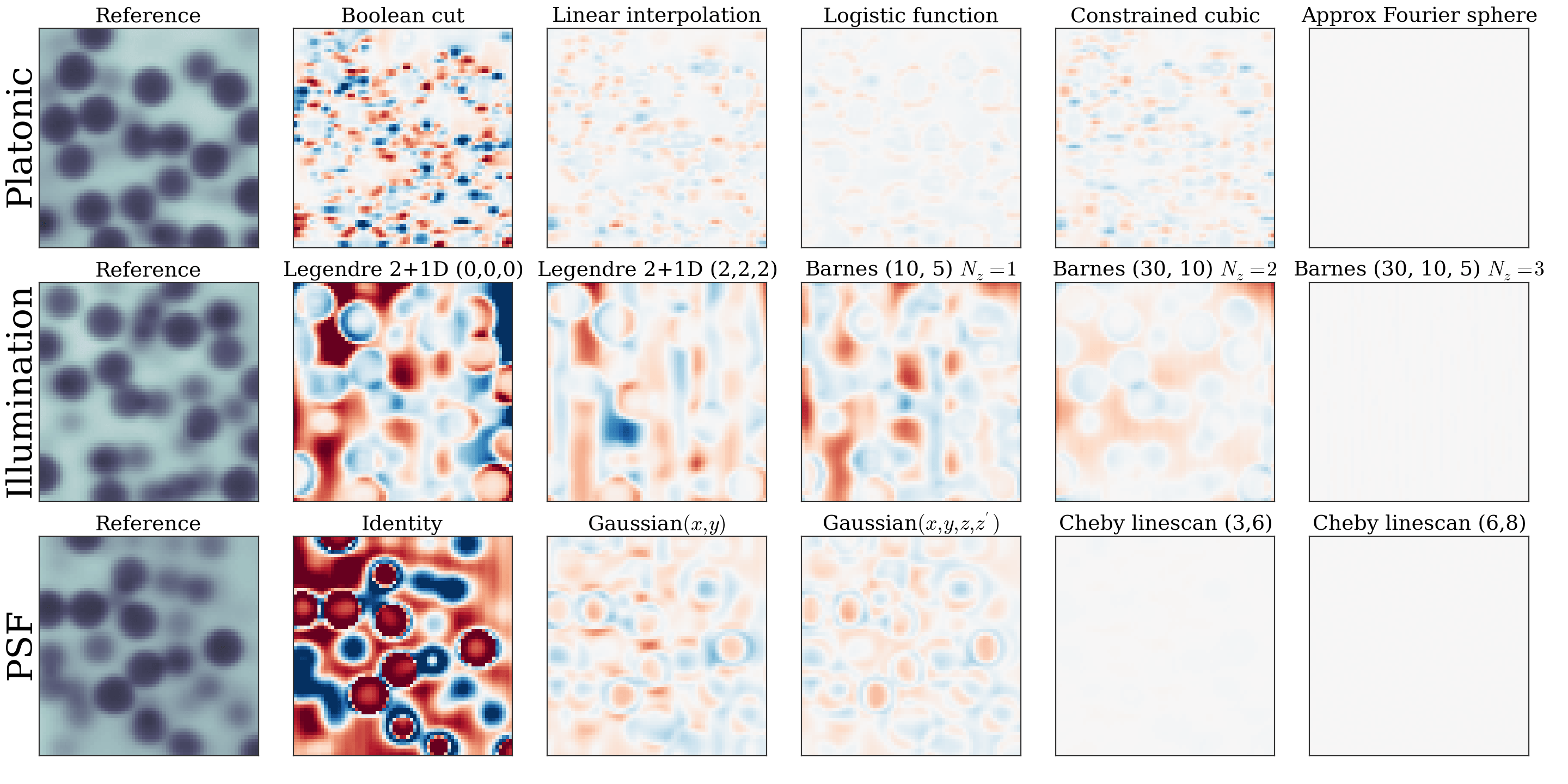}

\bcaption{Component complexity residuals}{.
Here we visually demonstrate the results of choosing between different
forms of model components as well as different parameterizations of a
single component.  We generate simulated microscope images using the model
components which we employ when fitting experimental data (left column) and
fit them with different choices of platonic image (top row), illumination
field (middle row), and point spread function (bottom row). Each choice is
labeled above its panel showing the residuals and each row is on a common
color scale. In the case of the platonic forms, the boolean cut, linear
interpolation, and constrained cubic display higher order multipole errors
while the logistic function's first correction is to the monopole (volume)
term (as shown by the presence of rings). In the illumination field,
stripes are present in the residuals until we use a Barnes interpolant with
$30$ control points. Past that, the ability to capture intensity as a
function of depth is the remaining term which we are able to fit with a
single extra Legendre polynomial in the $z$-direction. Finally, in the case
of the PSF, we see hard boundaries transitioning to softer boundaries using
a Gaussian PSF in both 3D and 3+1D. The residuals all but disappear when
the image is fit with our exact line-scan confocal PSF model
(Eq.~\ref{eq:PSFcalc}) approximated by a Chebyshev polynomial in 3+1D.
}

\label{fig:model_complexity}
\end{sidewaysfigure}

We would like to systematically investigate at what level omitting a detail of the image formation from the model affects the fitted parameters. We can understand this quantitatively by examining the optimization procedure. Let us assume that the true image formation is completely described by a set of $N$ parameters $\vec{\Theta}$. Then, near its maximum, the log-likelihood is approximately quadratic: $\log \mathcal{L} = \frac 1 2 \sum_{ij}H_{ij}\Theta_i \Theta_j$, where the true value of the parameters is arbitrarily set to $\vec{\Theta}=0$. Empirically, we find that with the starting parameter values provided by our initial featuring, the log-likelihood is extremely well-approximated by a quadratic.

If our model were complete, then the maximum of $\log \mathcal{L}$ would be exactly at the true parameter values $\vec{\Theta}=0$. However, our model is incomplete. This means that, instead of fitting all $N$ parameters $\vec{\Theta}$, we only fit the first (say) $M$ parameters, which for convenience we denote as $\vec{\theta}$. Thus we can write the log-likelihood as three separate terms:
\begin{equation}
\log \mathcal{L} = \frac 1 2 \sum_{i,j=1}^M H_{ij} \theta_i \theta_j + \sum_{i=M+1}^N \sum_{j=1}^M H_{ij}\Theta_i \theta_j + \frac 1 2 \sum_{i,j=M+1}^N H_{ij} \Theta_i \Theta_j \quad .
\end{equation}
The first term, containing only the parameters $\vec{\theta}$ that we are fitting, is the quadratic in the reduced space, with a maximum at the true parameter values. The unimportant third term reflects the separate contribution to $\log \mathcal{L}$ of the unknown or ignored portions of the model, and is constant in the $\vec{\theta}$ space. However, the second term mixes both the fitted parameters $\theta$ and the unknown parameters $\Theta_j$. This mixing results in a linear shift of $\log \mathcal{L}$ in the $\vec{\theta}$ space away from the true parameters, and causes a systematic bias due to an incomplete model. Minimizing $\log \mathcal{L}$ with respect to $\theta$ gives the fitted values of the parameters gives an equation for the best-fit incomplete model parameters $\vec{\theta}$:
\begin{equation} \label{eq:param_diff}
\theta_j = \sum_{k=1}^M {\bar{H}^{-1}}_{jk} \sum_{i=M+1}^N H_{ik}\Theta_i
\end{equation}
where $\bar{H}^{-1}$ is the inverse of the sub-block $\bar{H}$ of the Hessian matrix $H$ that corresponds to the fitted parameters $\vec{\theta}$.

We can use equation~\ref{eq:param_diff} to estimate the effect on one of the estimated parameters $\theta_j$, if we ignore one aspect of the generative model $\Theta_k$. Ignoring the off-diagonal terms in $H^{-1}$ to capture the scaling gives $\theta_j \approx H_{kj}\Theta_k / H_{jj}$. Thus, the error in the fitted parameter $\theta_j$ is proportional to both the coupling $H_{kj}$ between that parameter and the ignored aspect of the generative model, and the magnitude of the error of the generative model $\Theta_k$.

\subsection{Component complexities}

There are several choices one can make concerning the form and complexity of
each of the components of our model image. As discussed in the
Section~\ref{sec:generative_model}, we have implemented many forms of the
platonic image, illumination field, and point spread function and each one of
these forms has a varying number of parameters with which to fit. How do we
decide which form to use and at which complexity (number of parameters) to
stop? To decide on a per-image basis, we could employ Occam's factor, which is
a measure of the evidence that a model is correct given the
data~\cite{mackay2003information}. In practice, however, we are mainly
concerned with how these models influence the underlying observables
which we are attempting to extract. That is, we wish to use knowledge of the
physical system to check which model best predicts the particle locations and
sizes. To do so (as mentioned in the main manuscript), we often turn to
particle sizes versus time as well as particle overlaps, both physical
statements that assert almost no assumptions on our system.

We can also get a sense of the magnitude of the effect these choices have on
inferred positions and radii by creating synthetic data and fitting it using a
simpler model. In Fig.~\ref{fig:model_complexity} we show the residuals of such
fits for various simplifications made to the platonic form, illumination field,
and point spread function. In the left columns of the figure we see the
reference image formed using the most complex image model available and in each
row the residuals for each choice with a description of that choice above the
panel. For all but the last column, in which we fit the image with the exact
model once again, we can see systematic errors in the fit. We compute how much
these residuals influence the extracted positions and radii and report these
errors in Table~\ref{table:model-complexity}. In particular, most choices of
platonic image aside from the naive boolean cut do not influence particle
featuring below an SNR of $30$. However, the complexity of the illumination
field always matters until all long wavelength structure is removed from the
image. Finally, the choice of PSF is crucial, requiring the use of a calculated
confocal PSF to even approach the CRB.

{\renewcommand{\arraystretch}{1.5}
\begin{center}
\begin{table}
\begin{tabular}{c@{\hspace{1em}} | l | c | c |}
\cline{2-4}
& Fitting model type &
Position error (px) &
Radius error (px) \\ \hline \hline
\multirow{5}{*}{\rotatebox{90}{\textbf{Platonic form}}}
& Boolean cut              & $0.03376$ & $0.01577$ \\ \cline{2-4}
& Linear interpolation     & $0.00778$ & $0.00386$ \\ \cline{2-4}
& Logistic function        & $0.00411$ & $0.00352$ \\ \cline{2-4}
& Constrained cubic        & $0.00674$ & $0.00249$ \\ \cline{2-4}
& Approx Fourier sphere    & $0.00000$ & $0.00000$ \\ \cline{2-4}
\hline
\multirow{5}{*}{\rotatebox{90}{\textbf{Illumination}}}
& Legendre 2+1D (0,0,0)         & $0.18051$ & $0.13011$ \\ \cline{2-4}
& Legendre 2+1D (2,2,2)         & $0.06653$ & $0.03048$ \\ \cline{2-4}
& Barnes (10, 5) $N_z=1$        & $0.13056$ & $0.06997$ \\ \cline{2-4}
& Barnes (30, 10) $N_z=2$       & $0.04256$ & $0.02230$ \\ \cline{2-4}
& Barnes (30, 10, 5) $N_z=3$    & $0.00074$ & $0.00022$ \\ \cline{2-4}
\hline
\multirow{5}{*}{\rotatebox{90}{\textbf{PSF}}}
& Identity                        & $0.54427$ & $0.57199$ \\ \cline{2-4}
& Gaussian$(x,y)$                 & $0.47371$ & $0.14463$ \\ \cline{2-4}
& Gaussian$(x,y,z,z^{\prime})$    & $0.34448$ & $0.04327$ \\ \cline{2-4}
& Cheby linescan (3,6)            & $0.03081$ & $0.00729$ \\ \cline{2-4}
& Cheby linescan (6,8)            & $0.00000$ & $0.00000$ \\ \cline{2-4}
\hline
\end{tabular}
\bcaption{Position and radii errors by model complexity.}{
Here we tabulate the position and radius errors associated with the
model component choices made in Fig.~\ref{fig:model_complexity}. Note
that while the components with the largest impact on determining
underlying parameters are the ILM and PSF, the choice of platonic image
cannot be ignored in order to reach the theoretical maximum resolution.
Interestingly, in the case of PSF selection, Gaussian$(x,y,z,z^{\prime})$
(3+1D) is almost no better at extracting particle
positions than Gaussian$(x,y)$ (2D). However, its ability to
extract particle sizes increases by $3$ since it takes into
account the variation of the PSF in space. Additionally, in the case of
the ILM, capturing the stripes in the illumination using a $30$ control
point Barnes increases the resolution by $3$ whereas capturing
the illumination's dependence in depth causes the resolution to
increase $10$ fold.
}
\label{table:model-complexity}
\end{table}
\end{center}
}
\subsection{Scan jitter}

Confocal microscopes operate by taking an image with the lens at a fixed $z$
position to create one layer of the three-dimensional image, then moving the
lens up a fixed amount to take the next layer. In our generative model, we
assume that these steps of the lens (and the resultant image slices) are
perfectly equally spaced by an amount which is fitted internally. However, a
real confocal microscope will have some error in the vertical positioning of
the lens. As a result, the actual image taken will not be sampled at exactly
evenly spaced slices in $z$, but at slices that are slightly shifted by a
random amount.

\begin{figure*}

\includegraphics[width=0.95 \textwidth]{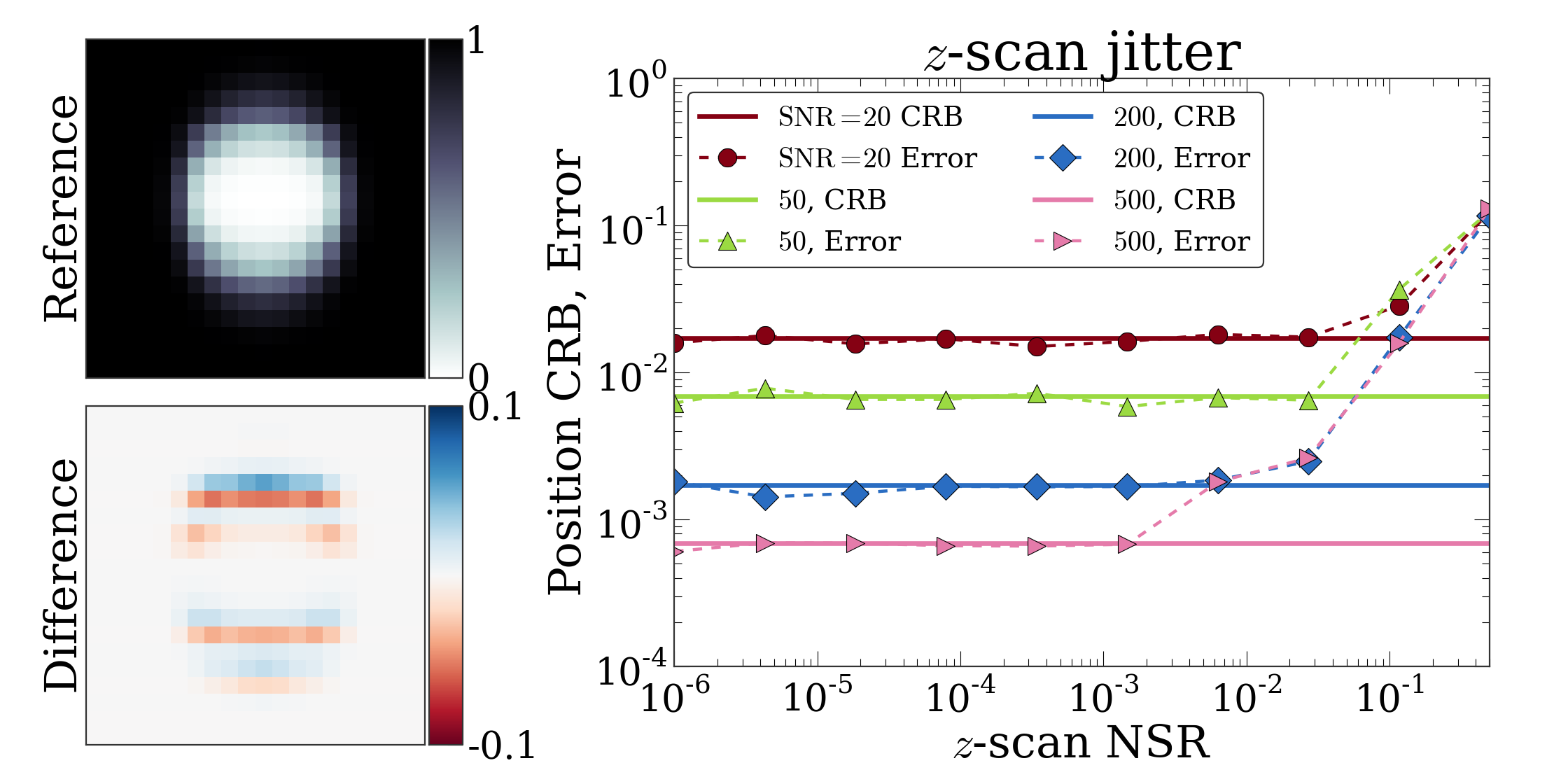}

\bcaption{Lens Positioning Jitter}{ (a) The $xz$ cross-section of a simulated image of a $5\px$ radius colloidal particle taken with a 10\% error in the lens positioning. (b) The difference between the image with positioning error and a reference image with zero positioning error. The differences between the images are both random and small, for this image no more than 7\% of the perfect image intensity. (c) The effect of lens positioning error on featured particle positions, at signal-to-noise ratios of 20, 50, 200, and 500. The solid symbols and dashed lines show the position error for images with imperfect lens positioning, while the solid lines denote the Cramer-Rao bound for an image with no positioning error. At lens positioning errors of $\approx 10\%$ or larger, the error in featured positions from the $z$-slice jitter dominates that from the simple image noise, even for an SNR of 20. However, the featuring error due to a $z$ jitter of $\approx 1\%$ is less than the error due to image noise, for any noise level than can be captured by an 8-bit camera.
}

\label{fig:z_jitter}
\end{figure*}

To test the effect of this $z$-scan jitter on our parameter estimation, we
simulate images taken by a confocal microscope with imperfect $z$-positioning.
Instead of sampling the image at a deterministic $z$ position, we instead
sampled the image at a $z$ position shifted from the ideal position by an
uncorrelated Gaussian amount of varying standard deviation. A representative
image of a $5\px$ radius particle with a step positioning error of 10\% is
shown in Fig.~\ref{fig:z_jitter}(a). There is very little difference between
this image with $z$ jitter and the perfectly-sampled image, as shown by the
difference image in panel~b. We then fit an ensemble of these images at varying
image SNR levels, over a random sampling of image noise, $z$-jitter noise, and
random shifts of particle positions by a fraction of a pixel.

The results of these fits are shown in Fig.~\ref{fig:z_jitter}c, showing the
actual error in the featured positions versus the size of the $z$-positioning
noise. For our confocal which is equipped with a hyper-fine $z$-positioning
piezo, we expect the $z$ positioning error to be a few $\nm$, or a few percent
of a pixel. For a 3\% error in positioning, the signal-to-noise ratio must be
$\approx 100$ for the effects of $z$-positioning jitter to be comparable to the
theoretical minimum effect from the image noise. This small effect of the error
is partially due to the large size of our particle. If each $z$ slice of the
image is randomly displaced with standard deviation $\sigma$, then we expect
roughly a $\sigma/\sqrt{N}$ scaling for the final error in the particle's
$z$-position, where $N$ is the number of $z$ slices the particle appears in. A
$5\px$ diameter particle with a $4\px$ axial point-spread function occupies
$\approx 18$ difference slices, decreasing the effect of scan noise by a factor
of $\approx 4$ and putting it below the CRB for our data.

As the error in $z$-positioning increases, however, the effect on the featured
particle positions increases correspondingly. The error due to a $\approx 10\%$
$z$ jitter is comparable to the CRB for image noises of $SNR=20$. For
exceptionally large $z$-jitters of $40\%$ the error due to the lens positioning
dominates all other sources of error. However, even with this large error in
lens positioning, the error in featured positions is still only 10\% of a
pixel, or about 10 $\nm$ in physical units.

\subsection{Missing and Edge particles}

The point spread function delocalizes the particle's image over a region larger than the particle's size. As a result, if two particles are close enough together, their images can overlap. This overlapping is a significant problem for heuristics such as centroid fitting, as the true particle centers do not coincide with the fitted centroid. In contrast, PERI's accuracy is negligibly affected by the presence of a second, close particle, since PERI correctly incorporates close particles in its generative model. The CRB of two touching, $5\px$ diameter particles increases by only $\approx 3\%$, and PERI finds particles to this accuracy when close.

\begin{figure*} [tp]
\includegraphics[width=0.95 \textwidth]{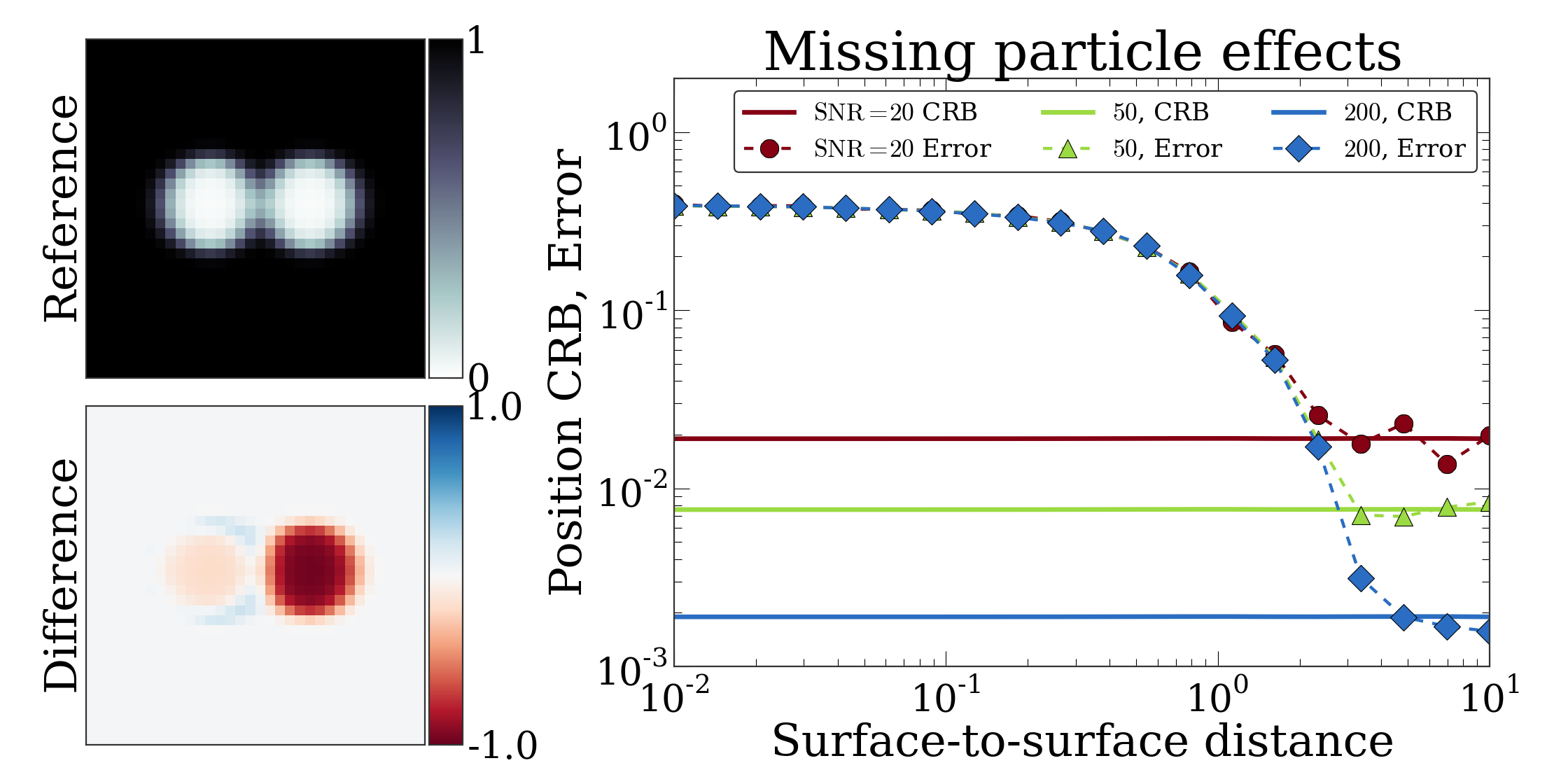}

\bcaption{Effect of missing particles}{.
(a) The $xz$-cross section of an image of two $5 \px$ radius particles placed in contact. (b) The
difference image for a bad generative model that includes only the particle on the left. To minimize
the effect of the missing right particle, the left particle is drawn to the right and expanded in
radius. This effect is visible as the red and blue ring on the right border of the left particle.
(c) The error in position along the separation axis, as a function of true surface-to-surface distance,
for a model with a missing particle. When the particles are separated by $\approx 10 \px$ the featured
particle is located correctly. However, as the particles get closer than $\approx 2\px$ significant
biases start to appear. These biases saturate at a separation of $\approx 0.1 \px$, corresponding to
a featuring error of $\approx 0.4 \px$.
}

\label{fig:missing_particle}
\end{figure*}

However, large systematic errors can affect PERI when one of these particles is missing in the generative model. This situation is illustrated in its simplest form in Fig.~\ref{fig:missing_particle}. If one of the two touching particles is missing from the generative model, then the second particle will be enlarged and drawn into the first particle's void to compensate, as shown in panel~b. As a result, the missing second particle will severely bias the fitted positions and radii of the first particle. Figure~\ref{fig:missing_particle}c shows the magnitude of this effect. For particles separated by $1\px$ or less, significant biases on the order of $0.4\px$ appear in the identified particle's featured position. These biases matter at essentially all values of the SNR, only being comparable to the CRB for $\textrm{SNR} < 1$. As a result, it is essential for PERI to identify all the particles in the image to return accurate results. For this reason, we take extra precaution and thoroughly search the image for missing particles before fitting, as detailed in section~\ref{sec:implementation}.

The biases caused by missing particles appear whether or not the missing particle is located inside or outside the image. As a result, accurately locating edge particles requires identifying all their nearby particles, even ones that are outside the image! We attempt to solve this problem by padding the Platonic and model images and the ILM by a significant portion, and including this padded extra-image region in both the add/remove and relaxation portions of the PERI algorithm. Nevertheless, it is extremely difficult to locate all the particles outside the image, for obvious reasons. As such, there is the possibility for moderate systematic errors to enter for particles located at or near the edge.

\begin{figure*} [tp]
\includegraphics[width=0.65 \textwidth]{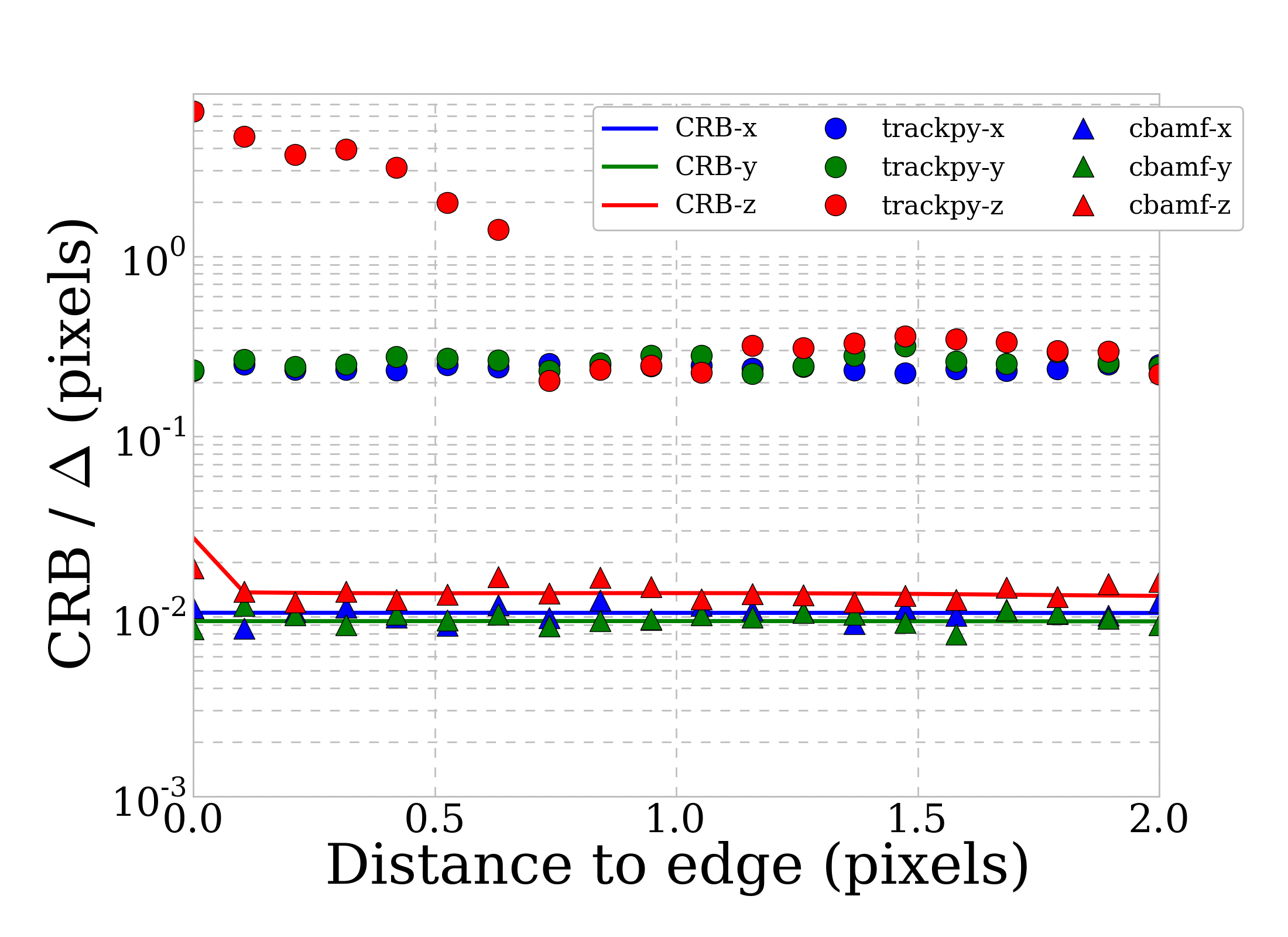}

\bcaption{Influence of particles outside of the image}{. Here we place one particle at $x=a$ and a
second particle at $x=-(a+\delta)$ so that one is completely inside the
image and the other outside. We plot the CRB for the $x$, $y$, and $z$
positions and radius $a$ of the interior particle as well as measured
errors for PERI in triangles and a centroid algorithm (trackpy~\cite{trackpy}) in circles as a function of the
position of the second particle.  When the exterior particle is further
than a pixel outside the image we see that the measurements of the interior
particle are constant.  However, as the PSF of the exterior particle begins
to overlap the interior particle the CRB and all measured errors increase
dramatically.  While PERI's measured error continues to follow the CRB,
trackpy's error increases beyond pixel resolution.  Note that pixel
separations at the edge are generic in colloidal images especially in dense
suspensions.
}

\label{fig:particles_outside_image}
\end{figure*}

Nevertheless, if the exterior particle is identified, PERI correctly locates the interior particle, as shown in Fig.~\ref{fig:particles_outside_image}. To demonstrate this, we create simulated images of two particles near the boundary of an image. One particle is placed at $z=a$ so that its edge just
touches the boundary while the other is placed at $z=-(a+\delta)$ on the other
side of the border.  We plot the CRB of the interior particle and the
measurement errors of both PERI and trackpy~\cite{trackpy} as a function of the exterior
particle's coordinate in Fig.~\ref{fig:particles_outside_image}. While the CRB
only changes by a factor of 2 as the particles come within contact, the
featuring errors grow drastically for traditional featuring methods due to
biases introduced by the exterior particle.  For this same data set, PERI
featuring errors follow the CRB allowing precise unbiased featuring of
particles at the edge of images.

\begin{figure*} [tp]
\includegraphics[width=0.65 \textwidth]{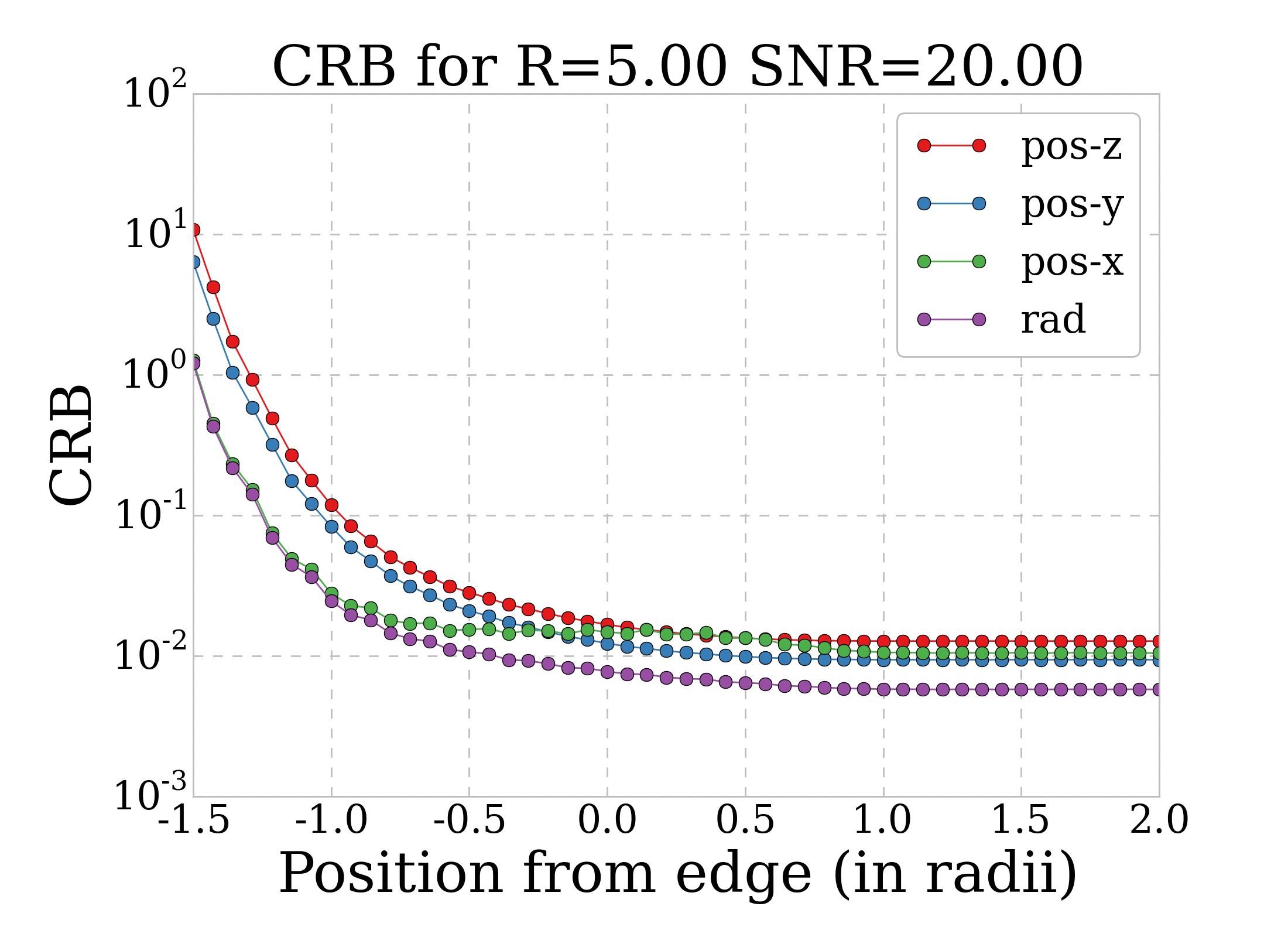}

\bcaption{CRB of edge particles}{. Here we calculate the \CRB~of the $x$,
$y$, and $z$ positions as well as radius (in red, blue, green, purple
respectively) for an isolated particle as a function of its distance to the
edge of the image.  For positive displacement (inside the image) we see
very little change with position as expected.  As parts of the PSF leak
out of the image (displacements close to zero, positive) we see that the
expected error increases slightly since information is lost.  Finally, as
the particle itself leaves the image, information is lost more dramatically
as indicated by a sharp rise in the CRB.  However, note that even at a
displacement of one radius $a$, the PSF allows us to locate the particle
outside of the image to within a pixel.  While in practice it is difficult
to identify these particles systematically, their presence can greatly
influence the measured positions of other edge particles.
}

\label{fig:crb_edge}
\end{figure*}

This apparent conundrum of edge particles presents an interesting positive side-effect. Missing edge particles affect the fits because they contribute a significant amount to the image. As such, we might expect that a particle outside the field of view can still be located very precisely. This prediction is borne out by a calculation of the \CRB, as shown in Fig.~\ref{fig:crb_edge}. Until the particle and PSF fall off the edge of the image (distance $> 1R$), the CRB remains constant for all particle parameters.  When the particle is centered on the image edge (distance of $0$), the
CRB is twice that of the bulk, intuitively corresponding to a loss of half of the
information about the particle.  As the volume of the particle leaves the
image, the CRB decreases as $1/\delta^2$ until the particle is no longer part
of the image.  Interestingly, Fig.~\ref{fig:crb_edge} shows that the PSF constrains the particle position to within $0.1\px$ even when the particle is entirely out of the image! If correctly seeded with a moderate guess for the particle position outside the image, PERI will locate the particle to a precision of the \CRB. However, in practice it is very difficult to seed these particles into PERI, as a slight change of the intensity at the image edge could be either a missing particle outside the image or a slight variation in the ILM near the image edge. Nevertheless, PERI is very good at locating particles that are partially outside the image.

\subsection{Pixel intensity integration}

Our generative model considers the image formed on the camera as if the camera pixels had an infinitesimal size. In reality, the camera pixels have a finite extent. As a result, the image at each pixel on the camera is not a discrete sampling of the light intensity, as in our generative model, but is instead an integration in the detector plane over the pixel's size.

\begin{figure*}

\includegraphics[width=0.95 \textwidth]{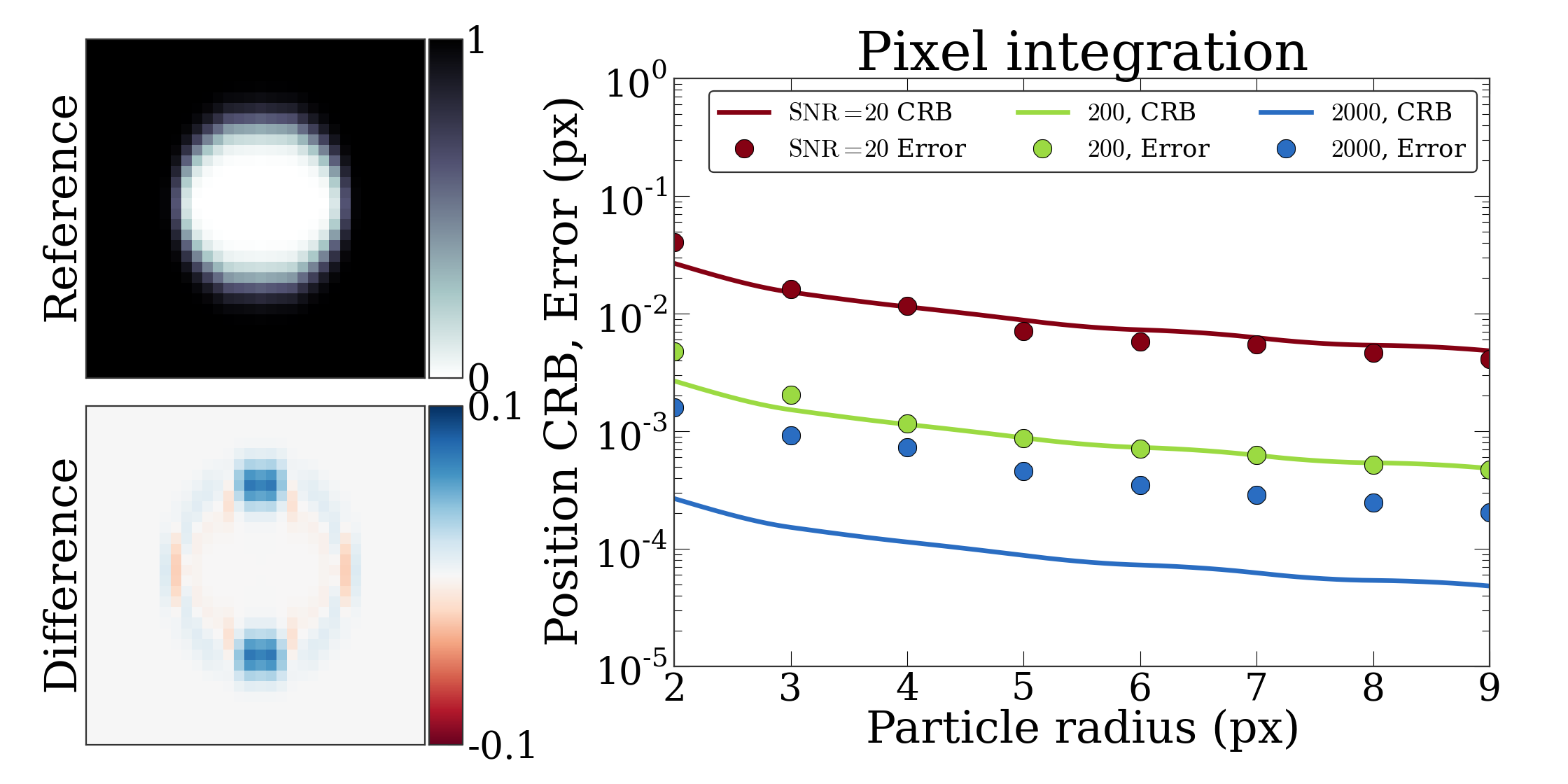}

\bcaption{Pixel Integration}{ (a) The $xz$ cross-section of a simulated image of a $5\px$ radius colloidal particle, where each pixel contains the light intensity integrated over its area instead of sampled at its center. (b) The difference between the pixel-integrated image and a reference image sampled at the center of the pixels. The differences between the images are small (10\%) and centered in a ring which has mean $0$ and is positioned at the particle's edge. (c) The effect of pixel integration on featured particle positions as a function of particle radius, at signal-to-noise ratios of 20, 200, and 2000. The solid symbols and dashed lines show the position error for images generated with pixel integration and fit without, while the solid lines denote the Cramer-Rao bound for the images (without pixel integration). Integrating over a pixel area has no effect on the featured positions for any SNR compatible with an 8-bit depth camera. The effect of pixel integration only starts to matter for an $\textrm{SNR} \ge 400$ (not shown).
}
\label{fig:pix_int}

\end{figure*}

To check whether the effect of pixel integration matters, we generated images that were up-sampled by a factor of $8$ in the $xy$-plane. We then numerically integrated these images over the size of each pixel. A representative image is shown in Fig.~\ref{fig:pix_int}a. There is very little difference between the $xy$-integrated image and the generative model, as visible in panel~b. We then fitted an ensemble of these $xy$-integrated pixel images, both over an ensemble of noise samples and over an ensemble of particle positions shifted by a random fraction of a pixel. The results are shown in Fig.~\ref{fig:pix_int}c. We find that there is no discernible effect of pixel integration at a SNR of $200$ or less. The error due to neglecting pixel integration becomes comparable to that due to noise only for $SNR \ge 400$, which is significantly higher than the maximum allowed by an ordinary 8-bit camera. Thus, the effect of integrating over a pixel size for a colloidal particle essentially always has a negligible effect on the fitted parameters.

\subsection{Diffusional motion}

A typical colloidal particle is not fixed in its location, but diffuses about
due to Brownian motion. For an isolated colloidal particle, this Brownian
motion results in a random walk with mean displacement $\langle \vec{x} \rangle
= \vec{0}$ and a mean-square displacement $\langle x^2 \rangle = 6Dt$ that is
linear in time, with a diffusion constant $D=kT/6\pi \eta R$ where $\eta$ is
the solvent viscosity and $R$ the particle radius. As a result, the microscope
takes an image not of a colloidal particle at a single position, but of an
integrated image of the colloidal particle over the trajectory that it has
diffused.

First, at what length- and time- scales is a colloidal particle de-localized
due to Brownian motion by a scale that is larger than the resolution? For a
$1\micron$ diameter particle in water to diffuse the 1 nm resolution provided by
PERI takes a fantastically small time of $t=1\nm^2/D \approx 10 \mu\textrm{s}$.
Even for our relatively viscous samples of $\approx 80\%$ glycerol and $20\%$
water this time slows down to only $\approx 600 \mu\textrm{s}$. These times are orders of magnitude faster than the $\approx 5 \textrm{ms}$ required by our confocal
to take a 3D image of the particle, corresponding to a 8 $\nm$ displacement.
Thus, a freely diffusing particle has always diffused much more than the
featuring errors than the uncertainty intrinsic to PERI.

However, this does not mean that the precision past $8\nm$ is empty. The
particle's positions are Gaussian distributed about its mean value during the
exposure time. While the extent of the distribution is much larger than the
PERI featuring errors, the particle's mean position during the exposure time is
well-defined. Moreover, the actual image on the camera from the diffusing
particle is a convolution of the particle's trajectory with a single particle
image. Since this convolution is like an averaging, we might expect that
the small Brownian excursions are averaged out in the image formation, and that
the image allows for accurate featuring of the particle's mean position.

We can use the formalism of Eq.~\ref{eq:param_diff} to show that Brownian motion does not affect our featuring accuracies. Let the particle's mean position be $\vec{\bar{x}}_0$, and its Brownian trajectory be $\vec{x}_0(t)$. Then the actual image $I(\vec{x})$on the detector is
\begin{equation}
I(\vec{x}) = \frac 1{t_\textrm{exp}} \int_0^{t_\textrm{exp}} I_0(\vec{x}_0(t))\, dt = I_0(\vec{\bar{x}}_0) + \frac 1 {t_\textrm{exp}} \int_0^{t_\textrm{exp}} I_0(\vec{x}_0(t))-I_0(\vec{\bar{x}}_0) \, dt
\end{equation}
where $I_0(\vec{x})$ is the image of one particle at position $\vec{x}$ and $t_\textrm{exp}$ is the camera exposure time. As before, we view the actual image as $I(\vec{x})=I_0(\vec{\bar{x}}_0;\,\theta) + (1-\Theta) \Delta I$, in terms of a group of fitted parameters $\vec{\theta}$ and an additional parameter $\Theta$ describing the effects of Brownian motion $\Delta I$. For the true image $\Theta=0$ but for our model image $\Theta=1$. Then equation~\ref{eq:param_diff} says the error will be $\theta_j \approx H_{kj}/H_{jj}$, where $H_{\Theta j} = \partial_\Theta \partial_{\theta_j} I= \partial_{\theta_j} \Delta I$. However, for small displacements the effect of Brownian motion on the image is
\begin{equation} \nonumber
\Delta I = \frac 1 {t_\textrm{exp}} \int_0^{t_\textrm{exp}} \frac {\partial I(\vec{\bar{x}}_0)}{\partial x_i} (\vec{x}-\vec{\bar{x}}_0) \, dt = 0
\end{equation}
since ${\partial I(\vec{\bar{x}}_0)}/{\partial x_i}$ does not depend on time. As a result, $\partial_{\theta_k}\partial_\Theta \Delta I = 0$ and there is no affect of Brownian motion on the image to first order in the displacements, \textit{i.e.} when the particle displacement is moderately small compared to the radius.

\begin{figure*}

\includegraphics[width=0.95 \textwidth]{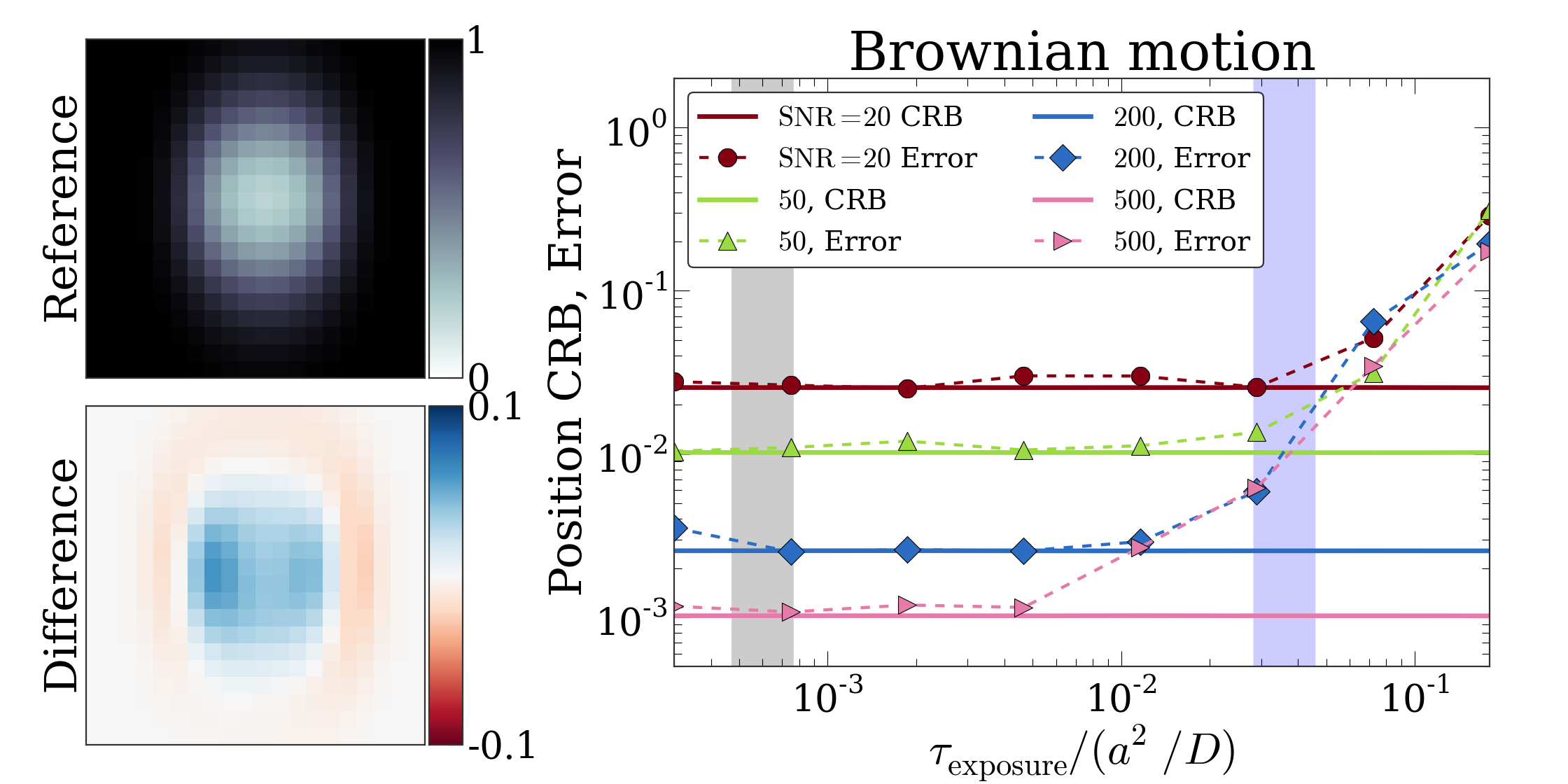}

\bcaption{Brownian Motion}{ (a) The $xz$ cross-section of a simulated image
of a $5\px$ radius colloidal particle undergoing strong Brownian motion
$\tau_\textrm{exposure} / (R^2 D)=0.01$ during the image formation. (b)
The difference between the diffusing-particle image and a reference
image without diffusion. The differences between the images are small
(10\%) and are mostly in a ring with mean $0$ at the particle's edge.
(c) The effect of Brownian motion on featured particle positions as a
function of exposure time, at signal-to-noise ratios of 20, 50, 200,
and 500. The image exposure time for our camera is located in the
shaded grey band for 20/80 water/glycerol and blue band for pure water. The
solid symbols and dashed lines show the error between the fitted
positions and the mean position in the particle's trajectory, while the
solid lines denote the Cramer-Rao bound for the generated images. At
our exposure times and SNR of 20, the effects of Brownian motion are
small compared to those from noise in the image.  Interestingly, for
higher SNR or slower exposure times, Brownian motion starts to have a
noticeable effect and must be incorporated into the image generation
model.

}
\label{fig:brownian_motion}
\end{figure*}

Finally, in Fig.~\ref{fig:brownian_motion} we show empirically that the effect
of Brownian motion is negligible for our exposure times. To create an image of
a diffusing particle captured by a slow camera, we simulated a 200 point
Brownian trajectory of a $R=5\px$ radius particle, generating an image for each
point in the particle's trajectory. We then took the average of these images as
the noise-free image captured by the microscope. One such image is shown in
Fig.~\ref{fig:brownian_motion}a. Once again, there is a slight difference
(10\%, as shown in panel~b) between the slow image of a diffusing particle and
the reference image taken of a particle at a single location. We then fitted an
ensemble of these images, over a variety of both Brownian trajectories and
noise samples. Figure~\ref{fig:brownian_motion}c shows the results of these
fits as a function of the mean displacement during the collection
$\tau_\textrm{exposure} / (R^2 D)$, where $\tau_\textrm{exposure}$ is the
exposure time of the camera and $D$ the particle's diffusion constant. Brownian
motion has a negligible effect on the featured positions for our experimental
images of freely-diffusing particles (camera exposure time of $100 \ms$ and
$D=0.007 \micron^2/\rm{s}$ corresponding to a $1\micron$ particle in 80:20
glycerol:water, corresponding to $\tau_\textrm{exposure} / (R^2 D)\approx
10^{-3}$). Interestingly, however, to achieve a higher localization accuracy at
a higher SNR of $\approx 200$, Brownian motion must be correctly taken into
account in the image formation. Incorporating Brownian motion at these high
signal-to-noise ratios would allow the teasing out of information about the
particle's trajectory from a single image.

\section{Implementation} \label{sec:implementation}

A typical confocal image is roughly $512 \by 512 \by 100$ pixels in size and
contains $10^4$ particles meaning that the number of degrees of freedom in our
fit is roughly $10^7$ described by $10^5$ parameters, a daunting space to
optimize.  On modern hardware using the highly optimized FFTW, the typical time
for an FFT the size of a single image is $\sim 1~\rm{sec}$.  Given this time, a
single sweep through all parameters would take an entire week while a full
optimization would consume a year of computer time.  However, since particles
have finite size, we are able to optimize most of these parameters locally with
a small coupling to global parameters (ILM, PSF).  Additionally, the finite 
intensity resolution of microscope sensors, typically 8 or 16 bits, allows us
to make further simplifications to our model. Here we describe the practical
algorithmic optimizations that we have made as well as the optimization
schedule that we have devised to quickly reach the best fit model.

\subsection{Partial image updates}

First and foremost, we optimize our fitting procedure by working in image
updates and only updating parts of the image that are required at any one time.
In order to modify the position of one particle by a small amount, the number
of pixels that are affected is simply $(2a + w)^3$ where $a$ is the particle
radius and $w$ is the PSF width, both in pixels.  For a typical particle, the
ratio of this volume to the entire image volume is typically $10^{-2}$ which
represents a speed up of the same factor due to the roughly linear scaling of
FFT performance with problem size ($N\log{N}$).  In addition, since the PSF decreases with distance from a particle's center, a localized object produces only a weak signal in regions far away from it. For confocal microscope PSFs, the distance scale associated with this
signal change is only a few tens of pixels.  Therefore, we employ a technique common
applied to inter-atomic potentials in molecular dynamic simulation -- we simply
cutoff the PSF at this distance scale allowing for exact partial updates.
By cutting off the PSF, we are able to incrementally apply image updates in an
exact procedure (up to floating point errors).  For example, when moving a
single particle from $\vec{x}_0$ to $\vec{x}_1$, we must simply calculate the
local image change given by
\begin{equation}
\Delta \mathcal{M}(\vec{x}) = \int \diff^3x^{\prime}\,\,[I(\vec{x^{\prime}})(1-c)(\Pi(\vec{x};\vec{x}_1) - \Pi(\vec{x};\vec{x}_0))] P(\vec{x}-\vec{x}^{\prime}; \vec{x}) \quad ,
\end{equation}
cf.~equation~\ref{eq:model}, then calculate $\mathcal{M} + \Delta\mathcal{M}$ only in a small local region around
the particle being updated.  We are able to use similar update rules for all
variables except for those effecting the entire image such as the PSF, offset,
$\zscale$, and estimate of the SNR.

Additionally, in our code, we generously employ the principle of ``space-time
trade-off'' in which we cache intermediate results of all model components and
reuse them later in the computation.  In particular, we maintain a full
platonic image and illumination field, which we update along with the model
image. We also cache the calculated PSF so that we may utilize previous
results until the PSF is sampled.  In doing so, we are limited in our current
implementation by the speed of the FFT, which takes 70\% of the total runtime.

\subsection{Optimization of parameters and sampling for error}

Once an approximate initial guess is obtained by more traditional featuring
methods~\cite{crocker1996methods}, we optimize the parameters by fitting using
a modified Levenberg-Marquardt routine. Our Levenberg-Marquardt algorithm uses
previously-reported optimization strategies designed for large parameter spaces~\cite{Transtrum2012}. However, a Levenberg-Marquardt minimization requires the
matrix $J_{i\alpha} \equiv \partial m(x_i)/\partial \theta_\alpha$, which is
the gradient of each pixel in the model with respect to all the parameters. For
the $\approx 10^5$ parameters and $10^7$ pixels in our image, this matrix would
be many thousand times too large to store in memory. Instead, we construct a
random approximation to $J_{i\alpha}$ by using a random sub-section of pixels
$x_i$ in the image to compute $J$. This approach works well for the global
parameters (PSF, ILM, etc) but fails for the particles, which appear in a
relatively small number of pixels. For the particles, we instead fit small
groups of adjacent particles using the full $J_{i\alpha}$ for the local region of affected
pixels. As the global parameters and particle parameters are coupled, we
iterate by optimizing first the globals, then the particles, and repeating
until the optimization has converged.

Once the model is optimized, we can employ two different methods to extract the
errors associated with each parameter. Since we calculated the gradients $J$
during the optimization procedure, we can use this to find the covariance
matrix $(J^TJ)^{-1}$ which gives the correlated sensitivities of each
parameter. In practice this is the faster method and yields accurate results
and, as such, is our method of choice. However, additionally, we may use Monte
Carlo sampling to estimate parameter errors. Our Monte Carlo sampler sweeps
over each parameter and updates the particle position, accepting or rejecting
based on the change in the log-likelihood of the model. We use slice sampling
to produce highly uncorrelated samples, allowing an excellent error estimate
from only a few sweeps. Our error sampling doubles as a check for convergence.
If the log-likelihood increases after sampling, then the optimization has not
converged and either more Monte Carlo sampling or more traditional optimization
is needed.  In practice, when desired we check with $\approx 5-10$ Monte Carlo
sweeps, and ensure that the log-likelihood remains the same or fluctuates by a
few times $\sqrt{N}$, where $N$ is the number of parameters in the model.

\subsection{Source code}

A complete implementation of this method is provided in a Python package called
\texttt{peri}, whose source can be found online~\footnote{Source code and tutorial at \url{http://www.lassp.cornell.edu/sethna/peri/index.html}} along
with extensive documentation about the particulars of its implementation.
Additionally, it is available at \url{PyPI.org}, the central repository for
Python packages outside of the standard library.

\section{Benchmarks of featuring algorithms} \label{sec:benchmarks}

\begin{figure} [tp]
\includegraphics[width=1.0 \columnwidth]{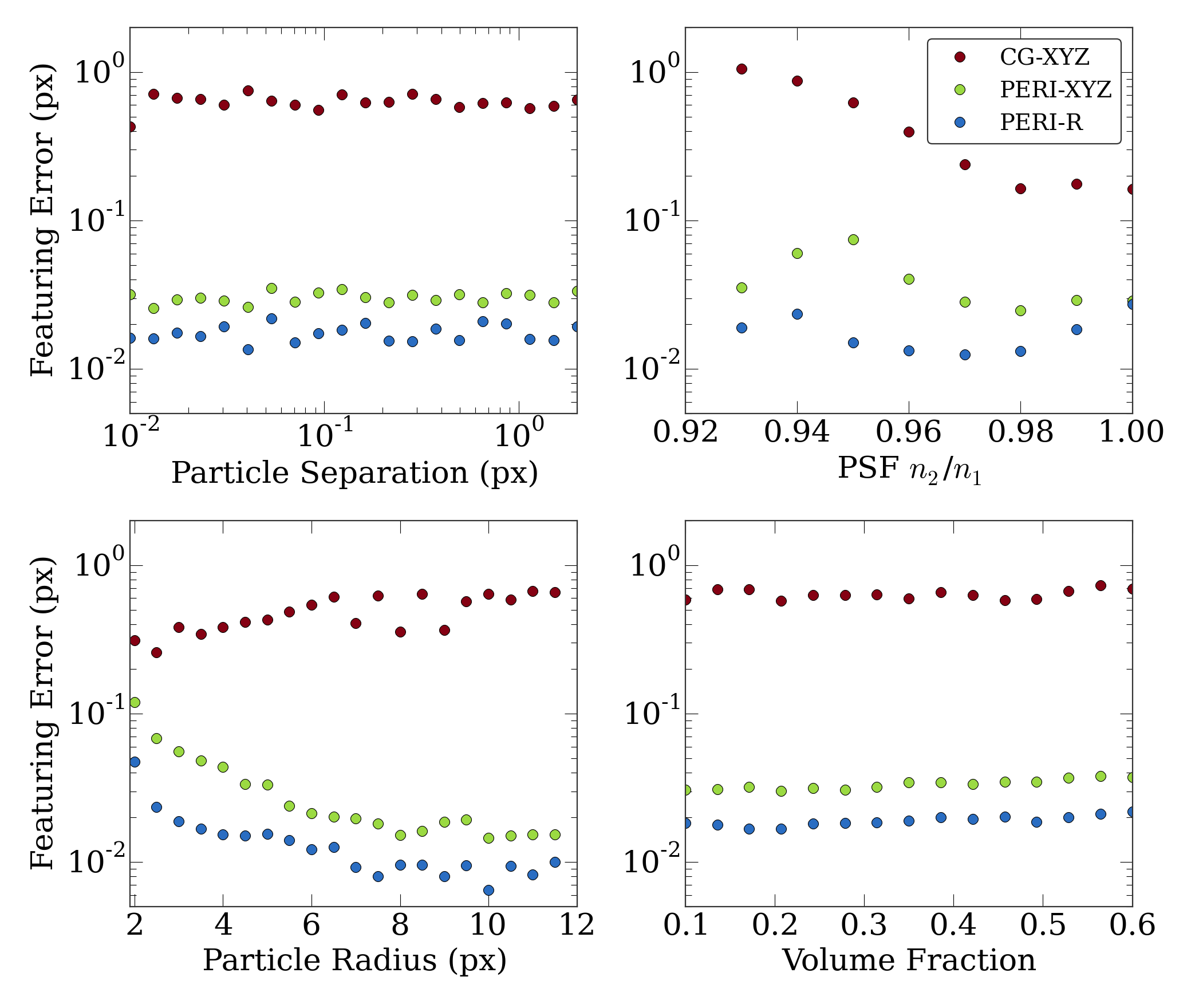}

\bcaption{Accuracy benchmark}{. We compare the featuring errors of PERI
and a traditional centroid (Crocker Grier or CG) featuring method with
the optimal featuring parameters. The panels show the featuring errors vs.
particle separation (upper left panel), vs PSF aberration through
the index mismatch $n_2/n_1$ (upper right panel), vs. particle radius (lower
left panel), and vs. the suspension volume fraction (lower right panel).
}

\label{fig:benchmark}
\end{figure}

\subsection{Generated Data}\label{subsec:gendata}
We check our algorithm by benchmarking it against physically realistic image models, as shown in Fig.~\ref{fig:benchmark}. For maximal realism, we generate these images with every model component in equation~\ref{eq:model} as realistic as possible. We use our exact calculation for line-scanning confocal microscopes, with physical parameters expected from an experiment. From the structure of our fitted line-scan confocal images, we re-create a random illumination field that closely mimics the power spectrum of our actual confocal. We position the particles randomly, without placing them preferentially on the center or edge of a pixel. Since real images have particles that are also outside or partially inside the image, we generate the image on a large region before cropping to an internal region, resulting in edge particles and particles outside the field of view.~\footnote{Unless otherwise specified, we use an index mismatch $n_2/n_1=0.95$, a ratio of fluorescent light to excitation light energies of $0.889$, an excitation wavelength of $488\nm$, and a lens aperture acceptance angle of $1.173$ corresponding to a $1.4$ NA lens. The particles are $1\micron$ in diameter, with a pixel size used of $100\nm$, and extend from a region from just above to $\approx 5 \micron$ above the coverslip.}

We then fit these algorithms both with PERI and with traditional centroid-based featuring algorithms. When we fit these images with PERI we start with initial guesses that are not near the correct parameter values, to ensure that our method is robust to realistic initial guesses. For the centroid featuring methods, there are several algorithms and variants that can be used. We use the most commonly used of these versions, as implemented by Crocker and Grier~\cite{crocker1996methods} in the IDL language. All of these centroid algorithms require the user to select various parameters, such as a filter size for smoothing of the noisy image and a mask size for finding the centroid positions. As is well-known in the colloid community, using the incorrect parameters can result in significantly poorer results. To overcome any possible limitation from using the incorrect parameters, we \textit{fit} all the possible parameters\footnote{We fit the $x,\,y,\,z$ bandpass sizes for both the lowpass and hipass filters, the centroid size or diameter, the particle mass size ``masscut'', the minimum particle separation, and a threshold below which pixels are ignored.} in the Crocker-Grier (CG) algorithm and use only the ones that produces the best global featuring of the data, as compared to the correct particle positions. (Centroid methods do not accurately find particle radii). Needless to say, an actual experimenter does not have access to the ground truth or to the optimal parameters for the featuring. Moreover, even with these optimal parameters, the centroid algorithm frequently misses a large fraction of particles, even in simple images. As such, we view the centroid featuring errors as unrealistically optimistic and probably not attainable with centroid methods even by experts. The results of these comparisons are shown in Fig.~\ref{fig:benchmark}.

When two particles are close, their images overlap due to the breadth of the point-spread function. This overlap causes centroid methods considerable difficulty. To compare the effects of PSF overlap on both PERI and CG featured positions, we generate an ensemble of realistic images with isolated pairs of particles at random orientations and at a fixed particle edge-to-edge separations. The upper-left panel shows these results for edge-to-edge separations from $0.01\px$ to $2.0\px$, with a fixed noise scale of about $0.05$ of the illumination amount. As the randomly-generated illumination fields vary from image to image, and the illumination varies from region to region within an image, there is not truly a global SNR for all of the images; the fluctuations in this SNR from image to image are the origins of the fluctuations in featuring error throughout Fig.~\ref{fig:benchmark}. PERI features particles at the Cramer-Rao bound regardless of their separation. In contrast, even at large separations of $2\px$, CG has significant errors due to particle overlaps.

Aberrations due to index mismatch significantly affect image quality and extracted particle locations. The upper right panel shows the effect of these aberrations on localizing isolated particles, as measured by the ratio between the index of refraction of the optics $n_1$ and of the sample $n_2$. Moving the ratio $n_2/n_1$ away from 1 increases aberration in the image. While increasing the aberrations in the lens negatively affects PERI's ability to feature particles, the localization accuracy always remains excellent. In contrast, CG methods perform poorly throughout, with extremely poor performance as the aberrations increase.

Since the CRB decreases with particle radius, we expect that increasing the particle radius should result in an increase in localization accuracy. The lower-left panel of Fig.~\ref{fig:benchmark} shows that PERI's precision improves with increasing particle radius. In contrast, the Crocker-Grier precision \textit{worsens} with increasing particle radius. We hypothesize this arises due to the flat intensity profile near the center of a large particle, whereas a centroid method assumes that the intensity is peaked at the particle center. As a result, slight noise can significantly worsen a large particle's localization with centroid methods. Conversely, centroid algorithms improve for small particles, performing only $3\times$ worse than PERI's localization accuracy for particles with radius $2\px$. For particles small to the PSF size, the image is essentially a single peak, which centroid methods work well for.

Realistic images taken with confocal microscopes consist of particles randomly distributed, occasionally close together and occasionally far apart. To examine the localization in these images, we use a Brownian dynamics simulation to create a random distribution of particles at volume fractions from $\phi=0.1$ to $\phi=0.6$. PERI localizes particle positions and radii excellently in all of these images, as visible in the lower-right panel. In contrast, centroid methods perform uniformly poorly, with localization accuracies of approximately half a pixel. Interestingly, these centroid algorithms do not localize significantly worse for dense suspensions despite the presence of more close particles, although they do frequently fail to identify particles.

Finally, we check how the complexity of our synthetic data affects the accuracy
of standard featuring methods. In Table~\ref{table:cg-complexity} we see,
surprisingly, that there is a non-monotonic relationship between positional
error and image complexity, becoming optimal when there is significant
striping in the image but little variation in depth. However, the rate of
missing particles decreases significantly with simpler models and rising to as
much as $40\%$ for our most complex model images. The effective resolution
of CG is never much smaller than a single pixel in these synthetic tests,
most likely due to pixel edge biases.

\begin{center}

\begin{table}
\begin{tabular}{| l | l | l | c | c |}
\hline
Polydispersity &
Illumination field &
Point spread function &
Position error &
\% Identified \\ \hline \hline
0.0   & Legendre 2+1D (0,0,0)           & Identity                        & 1.458 & 0.81 \\ \hline
0.0   & Legendre 2+1D (2,2,2)           & Gaussian$(x,y)$                 & 1.218 & 0.84 \\ \hline
0.01  & Barnes (10, 5), $N_z=1$         & Gaussian$(x,y,z,z^{\prime})$    & 1.015 & 0.75 \\ \hline
0.05  & Barnes (30, 10), $N_z=2$        & Cheby linescan (3,6)            & 0.819 & 0.64 \\ \hline
0.10  & Barnes (30, 10, 5), $N_z=3$     & Cheby linescan (6,8)            & 1.085 & 0.64 \\ \hline
\end{tabular}

\bcaption{Crocker-Grier featuring errors}{.
We show the effect of image complexity on position error and miss
rate for the CG featuring method using synthetic data. Surprisingly,
there is a non-monotonic behavior of error with complexity, hitting
a maximum for highly striped images that don't vary strongly with depth.
However, the featuring miss rate steadily decreases with complexity,
reaching only $60\%$ identified particles with our most realistic images.
}

\label{table:cg-complexity}

\end{table}
\end{center}

\begin{figure}
\includegraphics[width=1.0 \columnwidth]{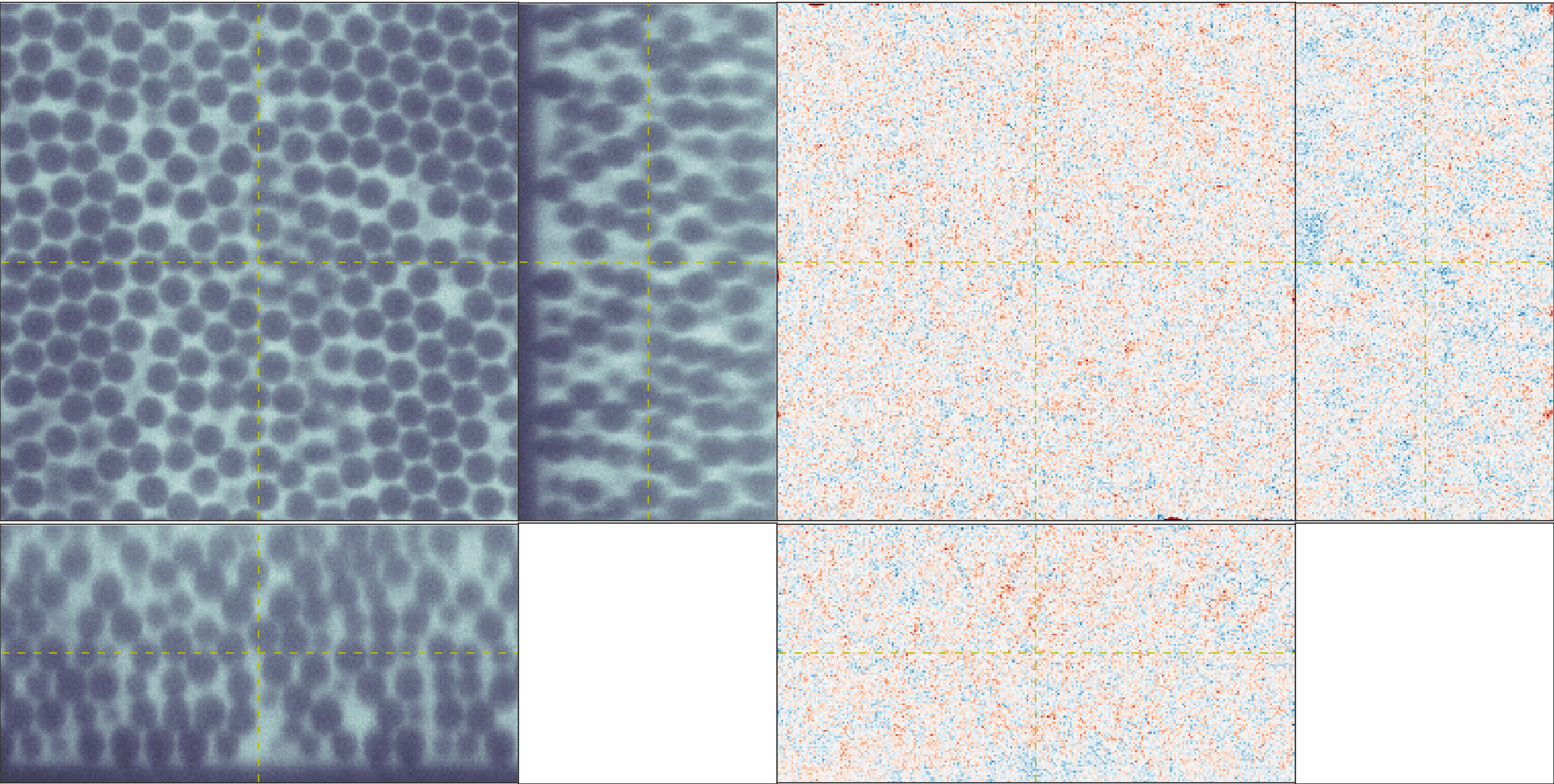}
\bcaption{Featuring Stuck Particles.}{
The raw image of the $2\micron$ sample of fixed particles (left), and the residuals to the fit (right), shown in $xy$, $yz$, and $xz$ cross-sections. Not only is the sample is extremely dense, but as the image is quite deep the index mismatch between the sample and the confocal optics creates strong aberrations deep into the sample. Despite these complications, PERI is able to fit this complex image and to accurately locate particles in it.
}
\label{fig:stuckparticles}
\end{figure}

\subsection{Fixed Particles}
Next, we check PERI on a sample of fixed particles. The sample is prepared by first making a dyed solution of $2\micron$ silica particles in an index-matching mixture of glycerol and water and loading the sample into a sample cell. At the edge of the sample we then add an equal amount of water-glycerol mixture saturated with salt (NaCl) and allow it to diffuse into the bulk of the sample over the course of a two weeks. As the salt diffuses in, it locally reduces the screening length and causes particles to strongly bind together. By letting the salt diffuse into the sample rather than mixing it in directly, the particles are able to sediment first before becoming fixed to each other, creating the dense sediment shown in figure~\ref{fig:stuckparticles}. We then image these particles with a five-second delay between images and analyze the resulting images using PERI. The particle positions fluctuate by $2.9 \nm$, $1.7\nm$, and $1.2\nm$ (median value) for $z$, $y$, and $x$, respectively, bounding the errors from above. (It is possible that some of the particles are not fixed to less than $2\nm$.) We find radii fluctuations of $0.8\nm$. 
﻿\section{Experimental Details} \label{sec:experimental}

To extract the interparticle potential, we use Molecular Dynamics simulations to find $\Psd$ and vary the parameters to find the best-fit $\Psd$. Since we know the particles' positions and radii via PERI, we seed the simulation with the featured particle positions and radii and relax the particle positions thoroughly before sampling for $\Psd$. Using the extracted particle parameters enforces both the correct amount of particle radii polydispersity and the number density of particles. In the simulation we use a standard DLVO potential, consisting of non-retarded van der Waals attractions and Debye-Huckel repulsion~\cite{russel1989colloidal}, augmented by gravitational settling. The free parameters we fit are the strength of the attraction, the strength and screening length of the repulsion, and the gravitational settling strength; physically these correspond to the Hamaker constant, a combination of the particle zeta potential and salt concentration, and the average particle density.

Since the $\Psd$ is measured from the simulation as a histogram with a finite number of samples, each simulated $\Psd$ is somewhat noisy. In light of this noise, we use a Nelder-Mead algorithm to find a good initial estimate of the fit parameters. We then refine this estimate of the fit parameters. First, we fit the ensemble of simulations to an approximate model which is locally linear in the fit parameters. We then use this linear model to estimate a new set of best-fit interaction parameters and refine our estimate of the potential; the curve plotted in Fig.~3 of the main text is the $\Psd$ generated from the linear model at the best fit parameters. To estimate uncertainties in the fit, we repeat this process by fitting a random subset of half of the simulations to a model and finding the new set of best-fit interactions. Repeating this 1000 times provides an estimate on the best-fit parameters as the mean of these best-fit parameters, and the uncertainty as the standard deviation of those parameters. Finally, we also obtain an estimate of systematic errors due to mis-featuring of particle positions and radii by fixing each particle's radius to be its mean value throughout all the images it is measured in. Surprisingly, fixing each particle's radius to a value that does not fluctuate in time worsens both the reconstrunction and the experimentally-measured $\Psd$, producing about three times as many overlaps. This probably arises because in some sense PERI directly measures the particle separations from the microscope image -- changing the separation of two particles slightly will considerably change the fraction of fluorescing dye separating them. Nevertheless, this fixed-radius data gives an order-of-magnitude estimate of any systematics in the experimentally-measured $\Psd$. In addition, we fit the inteparticle interactions for several different forms of the potential: hard spheres, electrostatic repulsion only, electrostatic repulsion and van der Waals attraction (DLVO theory), and DLVO theory combined with a short-ranged hydrophilic repulsion.

Table~\ref{table:dlvo-fits} shows the extracted potential parameters for all the interparticle interactions. Each interaction potential is fit two ways, by allowing the fitted particle's radius to vary with time and by fixing each individual particle's radius to its average value over the frames. With the exception of a pure hard-sphere potential, all of the various interaction potentials equally well-fit the data. In particular, fitting the data with just an exponentially-decaying electrostatic repulsion fits the data no better than including the van der Waals interaction. However, while our data does not exclude a nonzero Hamaker constant, the data is well-fit by Hamaker constants of a few kT. A hydrophilic repulsion is similarily not necessary to fit the data, but our data can accommodate hydrophilic repulsion of a reasonable strength and length scale. Since there are considerably more overlaps in the fixed radii data, we use the interaction potentials from the dataset with radii fitted by PERI as the best estimate of the fitting parameters, and the difference between the fits as an estimate of the systematic uncertainties from imperfect experimental data.

\begin{center}
\begin{table}

\begin{tabular}{|l|r|r|c|c|c|c|} \hline
& $U_{el}$, kT & $\lambda_{el}, \nm$ & $mg/\textrm{kT}, \nm$ & $A$, kT & $U_{hyd}$, kT & $\lambda_{hyd}$, $\nm$ \\
\hline \hline
Electrostatics, fitted $a$ & $100.6 \pm 3.4$ & $10.1\pm 0.06$ & $385 \pm 2$ & - & - & -
\\ \hline
Electrostatics, fixed $a$ & $67.8 \pm 2.3$ & $7.3 \pm 0.09$ & $378 \pm 4$ & - & - & -
\\ \hline
DLVO, fitted $a$ & $103.6 \pm 2.3$ & $10.4 \pm 0.05$ & $390 \pm 2$ & $0.231 \pm 0.006$ & - & -
\\ \hline
DLVO, fixed $a$ & $100.9 \pm 3.2$ & $7.5 \pm 0.06$ & $374 \pm 3$ & $0.286 \pm 0.010$ & - & -
\\ \hline
DLVO + Hyd., fitted $a$ & $121.5 \pm 0.7$ & $18.6 \pm 0.07$ & $376 \pm 2$& $0.496 \pm 0.004$ & $105.5 \pm 0.6$ & $4.7 \pm 0.03$
\\ \hline
DLVO + Hyd., fixed $a$ & $121.5 \pm 5.0$ & $25.2 \pm 1.8$ & $350 \pm 8$ & $0.513 \pm 0.02$ & $107.3 \pm 2.9$ & $4.9 \pm 0.1$ \\ \hline
\end{tabular}
\bcaption{Fitted Interaction Potentials}{.
The fitted interaction parameters for 3 sets of interparticle potentials, for
the positions and radii extracted from PERI (``fitted $a$'') and for each
particle's radius fixed to its mean value over the duration of many frames
(``fixed $a$''). The interparticle potentials are: Pure electrostatic repulsion
$U(\delta) = U_{el}e^{-\delta / \lambda_{el}}$, full DLVO theory (\textit{i.e.}
electrostatic repulsion and van der Waals attraction) with Hamaker constant $A$,
and full DLVO with an additional, short-ranged hydrophilic repulsion $U_{hyd}
e^{-\delta / \lambda_{hyd}}$. The uncertainties are the uncertainties in the fit
only.
}
\label{table:dlvo-fits}

\end{table}
\end{center}

\bibliography{Full_Bibliography}

\end{document}